\documentclass[aps,prx,twocolumn,superscriptaddress,longbibliography]{revtex4-2}
\usepackage{graphicx} 
\usepackage{amsfonts}
\usepackage{enumitem}
\usepackage[version=4,arrows=pgf-filled,textfontname=sffamily]{mhchem}
\usepackage{amsmath}
\usepackage{amssymb}
\usepackage{booktabs}
\usepackage{hyperref}

\newcommand{\varyingparamters}{$*$}
\newcommand{\mathd}{\mathrm{d}}

\begin{document}

\title{Enzyme-driven phase separation}

\author{Damiano Andreghetti}
\affiliation{Institute of Condensed Matter Physics and Complex Systems,
Department of Applied Science and Technology, Politecnico di Torino,
Corso Duca degli Abruzzi 24, 10129 Torino, Italy
}
\affiliation{Istituto Nazionale di Fisica Nucleare (INFN), Via Pietro Giuria, 1, 10125 Torino, Italy}
\author{Alfredo Braunstein}
\affiliation{Institute of Condensed Matter Physics and Complex Systems,
Department of Applied Science and Technology, Politecnico di Torino,
Corso Duca degli Abruzzi 24, 10129 Torino, Italy
}
\affiliation{Italian Institute for Genomic Medicine, Candiolo Cancer Institute,
Fondazione del Piemonte per l’Oncologia (FPO), Candiolo, 10060 Torino, Italy}
\affiliation{Istituto Nazionale di Fisica Nucleare (INFN), Via Pietro Giuria, 1, 10125 Torino, Italy}
\author{Luca Dall'Asta}
\affiliation{Institute of Condensed Matter Physics and Complex Systems,
Department of Applied Science and Technology, Politecnico di Torino,
Corso Duca degli Abruzzi 24, 10129 Torino, Italy
}
\affiliation{Istituto Nazionale di Fisica Nucleare (INFN), Via Pietro Giuria, 1, 10125 Torino, Italy}
\affiliation{Italian Institute for Genomic Medicine, Candiolo Cancer Institute,
Fondazione del Piemonte per l’Oncologia (FPO), Candiolo, 10060 Torino, Italy}
\author{Andrea Gamba}
\email{andrea.gamba@polito.it}
\affiliation{Institute of Condensed Matter Physics and Complex Systems,
Department of Applied Science and Technology, Politecnico di Torino,
Corso Duca degli Abruzzi 24, 10129 Torino, Italy
}
\affiliation{Istituto Nazionale di Fisica Nucleare (INFN), Via Pietro Giuria, 1, 10125 Torino, Italy}
\affiliation{Italian Institute for Genomic Medicine, Candiolo Cancer Institute,
Fondazione del Piemonte per l’Oncologia (FPO), Candiolo, 10060 Torino, Italy}

\begin{abstract}
The formation of polarized signaling domains on cell membranes is a fundamental example of biological pattern formation. While such patterns resemble structures from equilibrium phase separation, they are intrinsically nonequilibrium, driven by energy-consuming enzymatic cycles that switch molecules like phosphoinositides or small GTPases between distinct states. Here, we develop a minimal model of this enzyme-driven phase ordering process. Starting from microscopic reaction kinetics, we derive a mesoscopic theory that belongs to the class of active Model A with a global constraint. This framework yields an explicit mean-field phase diagram and closed-form expressions for key observables, such as interfacial tension, domain fractions, and phase coexistence boundaries, in terms of kinetic rates. In this context, phase coexistence is controlled by nonequilibrium parameters like catalytic rates and enzymatic asymmetry, rather than equilibrium parameters such as saturation concentrations. The resulting phase-separated domains rapidly exchange material with their surroundings. Their maintenance requires a continuous power input determined by enzymatic kinetics. The predicted phenomenology is consistent with experimental observations on reconstituted systems of phosphoinositide and Rab5 membrane patterning. We further study how metastable uniform states decay via nucleation of minority-phase domains and subsequent coarsening, driven by an effective interfacial tension. Using large deviation theory, we derive the critical nucleation radius under the action of the intrinsic, multiplicative chemical noise. The analytical results are quantitatively confirmed by stochastic simulations of the process. Our work provides a theoretical framework identifying key biochemical parameters controlling active phase separation on membrane scaffolds, offering testable predictions for experiments.
\end{abstract}

\maketitle

Eukaryotic cells display a remarkable degree of spatial organization, with a diverse array of supramolecular structures, from organelles to specialized membrane domains, that underpin essential functions like locomotion, signaling, and division~\cite{AHJ+22}.  Understanding the physical principles that guide the formation and maintenance of these dynamic structures remains a central challenge at the interface of biology and physics.
Phase separation has been proposed as a 
unifying concept, capable of explaining the localization of molecular factors and the formation of biological condensates~\cite{HWJ14,BLH+17}. 
Initially, this was done by drawing analogies from the physics
of 
quasi equilibrium, ``passive"
phase separation in polymer solutions and colloidal suspensions, where demixing is driven by a system's relaxation towards thermodynamic equilibrium, through the minimization of its free energy. 
However, biological self-organization is inherently nonequilibrium, often driven by energy-consuming, enzymatically controlled processes that are highly specific~\cite{FKL+19,Mus22,CN25}. 
These active mechanisms can lead to distinct phenomenologies, such as the suppression of coarsening and novel growth kinetics~\cite{ZHJ15,BWH+21,WZJ+19,YNW+22,BBW+25},
necessitating theoretical frameworks that go beyond classical equilibrium thermodynamics~\cite{BWH+21,YNW+22,ZHJ15,DPR16,CGA22,Rao+22,BWWF22,CN23,ZKZ23,DGMF23,ZMF25,CJ25}.

A canonical example of such active organization is the phase ordering of membrane-residing signaling molecules, such as phosphoinositides or small GTPases~\cite{HHL+19,LHH+21,CLS+20}, into signaling domains. Critically, these spatial patterns do not arise from passive equilibrium interactions but are driven by energy-dissipating enzymatic cycles. Feedback loops, whether direct or mediated by ancillary factors, promote the colocalization of enzymes with their reaction products, thereby establishing a self-reinforcing cycle. This leads to a phenomenon that resembles classical phase separation: the formation of distinct spatial domains enriched in a specific molecular state (e.g., GTP-bound or phosphorylated). However, the underlying mechanism is fundamentally different, as it is sustained by continuous, energy-consuming interconversion between states. Consequently, blocking the required energy supply (e.g., by removing ATP) leads to the dissolution of domains, demonstrating that the system is maintained far from thermodynamic equilibrium~\cite{HHL+19}.

In this work, 
we develop a minimal model of this enzyme-driven phase ordering process. We start from microscopic enzymatic kinetics and derive a mesoscopic theory that captures the system's essential features. The theory is characterized by a non-conserved order parameter coupled to a global constraint on molecular species fractions, which provides the stabilizing mechanism that makes phase coexistence possible~\cite{GCT+05,GKL+07,OIC+07,HF18,HBF18,BHF20}. The resulting dynamics is thus in the class of 
an
active 
version of
Model A with a globally conserved field~\cite{HH77,SM95}.
In this context, 
it is worth recalling that the mapping
of reaction-diffusion systems to Hohenberg-Halperin dynamical
universality classes~\cite{HH77} has deep roots. Already in the 1970s, the
Schl{\"o}gl model demonstrated how chemical reactions could produce
first-order phase transitions, controlled by an effective bistable
potential~{\cite{SCH72}}. Later, Goldbeter and Koshland highlighted the
emergence of switch-like behavior and bistability in systems governed by
antagonistic enzymes~{\cite{GK81}}. Such chemical reaction schemes typically
lack a locally conserved order parameter, as chemical factors can be freely
converted into each other. They thus tend to exhibit Model~A-type dynamics,
where phase-separated patterns disappear through coarsening (the progressive elimination of 
finer structures as larger ones grow)~{\cite{Bra94}}.
However, when reaction-diffusion systems are subject to global constraints,
stable coexistence of phases is possible. This principle is crucial in
mass-conserving reaction-diffusion 
theories of membrane polarization
where shuttling of enzymes between the membrane and a fast-diffusing cytosolic
reservoir induce Model~A dynamics with global conservation of an order
parameter: here, coarsening drives the system towards the stable
coexistence of competing phases~{\cite{GCT+05,GKL+07,GKL+09,SVN+12}}.
Importantly, in this scheme phase-separated domains emerge not only via
spontaneous linear instabilities of the Turing type, but also via the
potentially more controllable process of homogeneous or heterogeneous
nucleation from a uniform metastable
phase~{\cite{GCT+05,GKL+07,GKL+09,SVN+12,VGN+09}}. Similar 
models have
been proposed to explain the formation of polarity patterns in different
biological systems~\cite{OIC+07,Kru02,MK05,HMW03}, and are emerging as a general paradigm for
intracellular pattern formation {\cite{GKL+09,ZCT+15,HBF18,WBF23}}.
The study of the level sets of conserved quantities (reactive phase spaces)
and submanifolds of reactive equilibria (reactive nullclines)  
of 
mass-conserving
systems has unraveled nontrivial mechanisms of wavelength selection
and pattern stability in these nonequilibrium
settings~{\cite{HF18,BHF20,WBF23,RMS25}}. Depending on the structure of the
underlying reaction network, such systems can exhibit features of either
Model~A or Model~B dynamics, where Model~B is characterized by the added constraint of local mass conservation~{\cite{WBF23,RMS25}}. For systems that are microscopically active
due to energy consumption, coarse-graining is expected to yield ``active''
versions of the Hohenberg-Halperin classes. For instance, an active
modification of Model~B including terms that explicitly break detailed balance
has been successfully applied to the study of
motility-induced phase separation~{\cite{WTS+14,CT15,CN23,CN25}}. In general,
the derivation of mesoscopic theories from microscopic kinetics remains an
open challenge~{\cite{BLC+26}}. Here we address this point by showing
how underlying enzymatic cycles lead naturally to an active Model A dynamics
subject to a global constraint. In this framework, the nonequilibrium nature
of the system is captured through the kinetic non-linearities and the global
coupling to a reservoir, providing a direct link between biochemical energy
dissipation and stable phase coexistence.
This formulation, while incorporating realistic ingredients like local feedback and rapid 
exchange
with the reservoir,
remains analytically tractable. Within a mean-field approximation, we obtain an explicit phase diagram and closed-form expressions for key observables, including interfacial tension and phase-coexistence boundaries, directly in terms of basic kinetic rates. This allows us to identify how phase coexistence is promoted by enzyme affinity and catalytic asymmetries, and to determine the interfacial properties and energy dissipation required to sustain the active domains. The predictions of this analytical framework are confirmed by numerical simulations of a lattice-gas implementation of the full stochastic process. Moving beyond mean-field theory, we then analytically investigate fluctuation-driven events, deriving an explicit expression for the critical nucleation radius that quantifies the stochastic transition to a phase-separated state. Our results provide a theoretical foundation for interpreting experiments and identifying the key biochemical parameters that control active phase separation on membranes.

\section{Stochastic model}
\begin{figure}[b]
    \centering
\includegraphics[width=1\columnwidth]{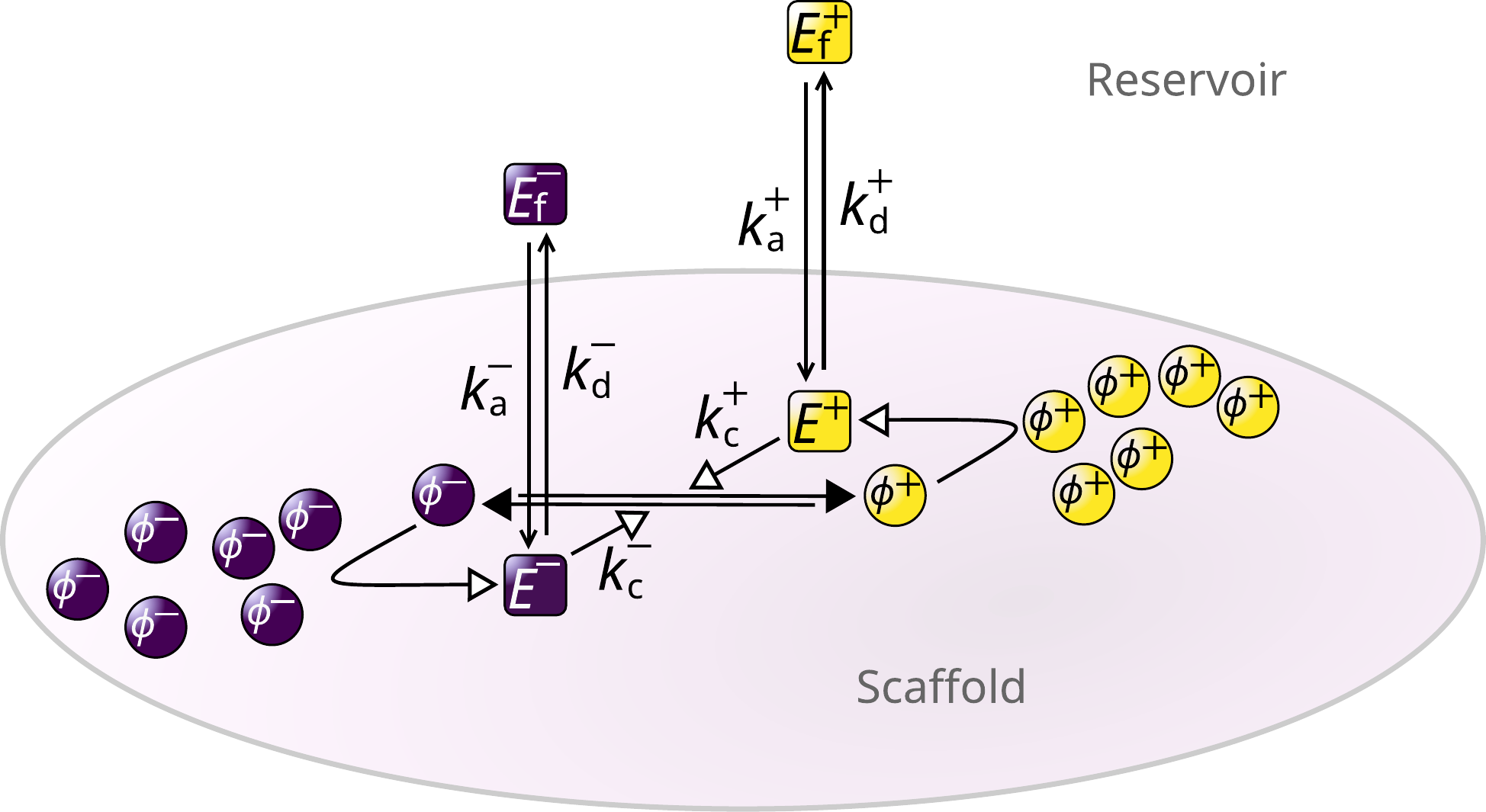}
    \caption{Schematic representation of the model 
    reaction network.
    The scaffold (e.g., a lipid membrane) is populated by two molecular species, $\phi^+$ and $\phi^-$, which can be converted into one another (horizontal arrows). This conversion process is catalyzed (oblique
    arrows) by scaffold-bound enzymes, $E^\pm$, with catalytic rates $k_\mathrm{c}^\pm$.
   Their free forms,
     $E_\mathrm{f}^\pm$, diffuse in a reservoir (e.g., the cytosol), binding to the scaffold with rates $k_\mathrm{a}^\pm$ and unbinding with rates~$k_\mathrm{d}^\pm$ (vertical arrows). The enzymes preferentially bind where the concentration of 
     their respective
     product molecules (i.e.,~$\phi^\pm$) is higher, 
     creating a self-reinforcing feedback loop
     (curved
     arrows).    
    See
    Tables~\ref{tab:fields} and~\ref{tab:parameters} of App.~\ref{app:adiabatic} for a summary of the relevant fields and parameters.
    }
    \label{fig:scheme}
\end{figure}
We consider a minimal model for bistable molecular switches on a slow-diffusion scaffold (typically, a membrane), inspired 
by 
canonical signaling systems
involving 
phosphoinositides
or
small
GTPases,  
where antagonistic enzymes  
regulate substrate states ~\cite{LBA01,MMS+12}.
The model 
(Fig.~\ref{fig:scheme})
comprises
two-state 
molecules 
$\phi^\pm$
that 
diffuse on 
the scaffold
with 
diffusivity~$D$.
Their
interconversion is catalyzed by enzymes $E^\pm$: 
\begin{equation}
    \ce{$E^\pm$ + $\phi^\mp$ ->[$k^\pm_\mathrm{c}$] $E^\pm$ + $\phi^\pm$} 
\label{eq:R1}
\end{equation}
We assume the enzymes operate 
via Michaelis-Menten kinetics
with
constants
$K_\mathrm{m}^\pm$,
thus allowing for both \hbox{linear}  and saturation regimes in the reaction rates.
The enzymes 
also diffuse on the scaffold with diffusivity $D$.
Furthermore, they shuttle between a fast-diffusing reservoir (typically, the cytosol, 
with diffusivity $D_\mathrm{f}\gg D$),
and the slow-diffusing scaffold,
preferentially binding to scaffold regions enriched in their 
corresponding
product~$\phi^\pm$:
\begin{subequations}
    \label{eq:R2}
    \begin{align}
        \ce{$E^\pm_{\mathrm{f}}$ + $\phi^\pm$ &->[ $k^\pm_\mathrm{a}$ ] $E^\pm$ + $\phi^\pm$} \\
        \ce{$E^\pm$ & ->[ $k^\pm_\mathrm{d}$ ] $E^\pm_{\mathrm{f}}$   }
    \end{align}    
\end{subequations}
Here, $E_\mathrm{f}^\pm$ and $E^\pm$ denote the free and scaffold-bound forms of the enzymes, respectively.
This feedback, where product recruits its own producer, provides a core mechanism for phase separation.
We also allow for a small, basal first-order conversion:
\begin{equation}
    \ce{$\phi^\mp$ ->[$k^\pm_\mathrm{b}$] $\phi^\pm$}
    \label{eq:R3}
\end{equation}
This reaction provides a background transition rate independent of enzymatic feedback.

The above described reaction-diffusion process involves multiple scales and a very large number of degrees of freedom (see App.~\ref{app:microscopic}).
As a consequence, 
integration of the corresponding master equation is unfeasible.
To make this problem tractable, we derived 
from the microscopic reaction rules, Eqs.~(\ref{eq:R1})--(\ref{eq:R3}),
mesoscopic
Langevin equations for a vector
$\vec{X} = (\phi^+, \phi^-, E^+, E^-, E^+_{\mathrm{f}}, E^-_{\mathrm{f}})$
of  
coarse-grained 
concentration fields, incorporating fluctuations
consistent with the central limit theorem (see App.~\ref{app:micromeso} and Ref.~~\cite{Gar85}):
$$\partial_t \vec{X}
(\vec{x},t)
=\vec{F}
[\vec{X}(\vec{x},t)
]+\vec{\Xi}
(\vec{x},t)
$$
 where $\vec{x}$ is a spatial coordinate,
 $\vec{F}$  
 is
 a deterministic drift term encapsulating 
 reaction  and diffusion kinetics,
 and~$\vec{\Xi}$ is  
 a stochastic
 noise term. 
  The force $\vec{F}$
is in general non-potential, and the system is 
still high-dimensional.
To reduce this system to a more manageable form, we
consider
the limit of infinite reservoir diffusivity ($D_{\mathrm{f}} \to
\infty$), 
effectively
treating the reservoir as a well-mixed, unstructured region where the concentrations $E_\mathrm{f}^\pm$ are spatially uniform.
Secondly, we 
assume 
that enzyme association and dissociation kinetics are much faster than molecular redistribution on the scaffold. 
On the characteristic timescale of phase separation, enzyme binding is thus effectively equilibrated, 
allowing for the adiabatic elimination of the enzyme degrees of freedom (see App.~\ref{app:adiabatic}).
In this regime,
the enzyme concentrations exhibit rapid stochastic variations, but their distribution is effectively slaved to the local substrate configuration, fluctuating around the local equilibrium:
\begin{equation}
\label{eq:enzymeequilibrium}
{K^\pm_\mathrm{d}}E^\pm 
=
\phi^\pm
E^\pm_{\mathrm{f}}
\end{equation}
where $K_\mathrm{d}^\pm=k_\mathrm{d}^\pm/k_\mathrm{a}^\pm$ 
(see App.~\ref{app:adiabatic}).
On the other hand, the
conservation of the total number
of each enzyme species 
imposes a
global constraint:
\begin{equation}
    \label{eq:stat_enz}
     \gamma^{-1}
    \langle 
    E^\pm\rangle_\mathrm{s} 
    +E^\pm_\mathrm{f}
    =
    E^\pm_\mathrm{tot}
\end{equation}
where 
$\langle\cdots\rangle_\mathrm{s}$ 
denotes 
the spatial average
over the scaffold, and
$\gamma$ is the ratio between the reservoir and scaffold
geometric
capacities (in the  
case of a three-dimensional cytosolic reservoir
and a two-dimensional membrane scaffold, this would correspond 
to 
their 
respective
volume-to-area ratio).
Together, 
Eqs.
~(\ref{eq:enzymeequilibrium})
and~(\ref{eq:stat_enz}) 
allow us to
determine~$E^\pm_\mathrm{f}$
from~$\langle
\phi^\pm
\rangle_\mathrm{s}$.

It is finally convenient to change variables to
the concentration difference $\phi$ and the
total concentration $c$ defined~as:
\begin{subequations}
    \begin{align}
        \phi(\vec{x},t) &\;=\; \phi^+(\vec{x},t) - \phi^-(\vec{x},t)
        \label{eq:phi-def}
        \\
         c(\vec{x},t) &\;=\; \phi^+(\vec{x},t) + \phi^-(\vec{x},t)
    \end{align}    
\end{subequations}
Notably, the total
concentration $c$ is unaffected by interconversion reactions and purely diffusive; it
thus exhibits random fluctuations around a homogeneous value
determined by the initial conditions (see App.~\ref{app:micromeso}).
Neglecting the diffusive 
fluctuations in 
this homogeneous field
(see App.~\ref{app:micromeso})
allows
 the system dynamics to be reexpressed in terms of 
 the 
 single,
 locally
 non-conserved
 order parameter $\phi$
 defined by Eq.~(\ref{eq:phi-def}).
 This order parameter is
subject to the constraint $-c\leqslant\phi\leqslant c$
and evolves according to the following
stochastic equation, 
which, due to the non-anticipating nature of chemical reaction events, should be interpreted in the It\^{o} sense (see App.~\ref{app:micromeso} and Ref.~\cite{Gar85}):
\begin{align}
    \label{eq:phi}
\partial_t\phi
(\vec{x},t)
&=D\Delta\phi
(\vec{x},t)
+A
[\phi]
+\sqrt{B
[\phi]
}\,\xi
(\vec{x},t)
\end{align}
where  $\xi
(\vec{x},t)
$ is a zero-mean Gaussian white noise with: 
\begin{align*}
    \langle \xi(\vec{x},t)\xi(\vec{x}',t')\rangle&=\delta(\vec{x}-\vec{x}')\delta(t-t'),
\end{align*}
and 
the 
drift and noise 
terms
are:
\begin{align}
    A
    [\phi]
    &=2(R^+
    [\phi]
    -R^-
    [\phi]
    )
    \label{eq:drift}
    \\
    B
    [\phi]
    &=4(R^+
    [\phi]
    +R^-
    [\phi]
    ) 
    \label{eq:noiseB}
\end{align}
with
$\phi^\mp\rightarrow \phi^\pm$ conversion rates $R^{\pm}[\phi]$, 
incorporating both Michaelis-Menten enzymatic kinetics and basal first-order transitions,
given by:
\begin{align}
R^\pm[\phi]&=
\frac{k^\pm_\mathrm{e}[\phi]}{2} \frac{c^2-\phi^2}{2\gamma K^\pm_\mathrm{m} +c\mp \phi}
+\frac{k^\pm_\mathrm{b}}{2}(c\mp\phi)
\label{eq:reactionrates}
\end{align}
and effective enzymatic rates  $k^{\pm}_{\mathrm{e}}[\phi]$
accounting for enzyme
depletion via scaffold recruitment, with:
\begin{align}
    k^\pm_\mathrm{e}[\phi] &= \frac{2k^\pm_\mathrm{c} \gamma E^\pm_\mathrm{tot}}{2\gamma K^\pm_\mathrm{d}+c \pm \langle \phi \rangle_\mathrm{s}} 
    \label{eq:alphas} 
\end{align}
The
multiplicative
noise 
$\sqrt{B
[\phi]
}\,\xi
(\vec{x},t)
$
in Eq.~(\ref{eq:phi})
is intrinsic, 
arising from the stochasticity of 
chemical reactions.
Fast enzyme shuttling brings
a correction of order $\epsilon=k_\mathrm{c}/k_\mathrm{d}$
to the 
noise amplitude $B[\phi]$ in Eq.~(\ref{eq:noiseB}) 
[see App.~\ref{app:adiabatic} for the complete expression, Eq.~(\ref{eq:Bafter_enslavement})].
This 
correction is subdominant 
in the fast-exchange (adiabatic) regime where the binding/unbinding of enzymes is much faster than catalysis, and
will 
be
neglected here
to maintain analytical tractability.
The drift and noise terms 
[Eqs.~(\ref{eq:drift}) and~(\ref{eq:noiseB})]
are closely related, due to the Poissonian nature of the underlying chemical events. 

The quadratic dependence on $\phi$ in the numerator of Eq.~(\ref{eq:reactionrates})
is the signature of a
local positive feedback. 
After 
adiabatic elimination of the enzyme degrees of freedom,
this feedback arises
from the underlying
structure of the reaction network: enzymes are recruited to
their own reaction products,
generating
an
effective pairwise attraction
between scaffold-bound molecules.
It is worth noting here that
nonlinearities
emerging
from both positive feedbacks
and the Michaelis-Menten saturation mechanism
can give rise to bistability~\cite{GK81,MJK08,HHL+19}. 

The dependence 
of the effective enzymatic rates  $k^{\pm}_{\mathrm{e}}[\phi]$
on the order parameter $\phi$ through its average $\langle\phi\rangle_\mathrm{s}$ 
in Eq.~(\ref{eq:alphas})
encodes the effects of a global negative feedback: since the total enzyme count is finite, the sequestration of enzymes within  domains
enriched
in one of the enzyme species
depletes the available pool in the reservoir.
This
gives
Eq.~(\ref{eq:phi})
the form of a time-dependent
Ginzburg-Landau equation
for a Model~A
system~\cite{HH77} constrained by a global conservation law, 
which arises from the finite, shared pool of enzymes being rapidly exchanged across the system via the reservoir.
The interplay between local self-reinforcement and global resource constraints constitutes the primary physical mechanism driving the appearance and stable coexistence of phase-separated domains~\cite{GCT+05,GKL+07,OIC+07,HF18,HBF18,BHF20}.

A summary of the fields and parameters in the model is
provided for convenience in  Tables~\ref{tab:fields} and~\ref{tab:parameters} of App.~\ref{app:adiabatic}.

The full dynamics of Eqs. (\ref{eq:phi})--(\ref{eq:alphas}) is still not analytically tractable. In the following, we exploit the fact that~$\langle\phi\rangle_\mathrm{s}$ is a slow variable compared to the local field dynamics. We thus analyze the system adiabatically, treating~$\langle\phi\rangle_\mathrm{s}$ as a fixed parameter, and 
later enforce the global constraint self-consistently 
for the determination of
the steady states.

The stochastic dynamics of Eq.~(\ref{eq:phi}) 
can be formulated
within the 
Martin-Siggia-Rose-Janssen-De\,Dominicis  
path-integral framework~\cite{MSR73,Kam23}.
In this context, the stochastic evolution is mapped onto a path integral over the field 
$\phi$
and an auxiliary response field
$\tilde\phi$
encoding noise realizations. This construction naturally introduces a generating functional, whose logarithm plays a role analogous to 
a free energy, as it generates correlation and response functions, while its extrema single out the most probable fluctuating trajectories~\cite{MSR73,Kam23}. This
functional
is 
explicitly
constructed by averaging over noise
distributed according 
to
$P[\xi]\propto \exp\left(-\frac{1}{2}\int\mathrm{d}t\int \mathrm{d}\vec{x}\,\,\xi^2\right)$, 
while enforcing the Langevin equation as a constraint:
\begin{eqnarray}
      && \int D\xi \,P[\xi] \int D\phi\, 
      {J}[\phi]\,
      \delta (\partial_t\phi-D\nabla^2\phi-A -
      \sqrt{
    B}
    \, \xi) \nonumber
\end{eqnarray} 
where the  
Jacobian determinant
$
{J}$
of the argument of the delta function is unity in the
It\^o regularization, 
which is the natural choice for intrinsic noise.
Representing the delta function by means of the auxiliary response field~$\tilde\phi$ 
and integrating out the noise,
 the generating functional can be recast as~\cite{Kam23}:
\begin{eqnarray}
    && \int \!D\phi\! \int \!D\tilde{\phi} \ \mathrm{e}^{-S[\phi,\tilde{\phi}]}  \nonumber
\end{eqnarray} 
with 
the action $S$ and 
Hamiltonian $H$ defined by:
\begin{subequations}
    \begin{align}
        S[\phi,\tilde{\phi}]&=
     \int
    \mathrm{d}t
    \int 
    \mathrm{d}\vec{x}
    \left\{ \tilde{\phi}
    \,[\partial_t\phi-D\nabla^2\phi-A]-\frac{1}{2}
    B
    \tilde{\phi}^2 \right\}\nonumber\\
    &=\int\mathrm{d}t\left\{ \int \mathrm{d}\vec{x}\,\tilde{\phi}\,\partial_t\phi-H[\phi,\tilde{\phi}]\right\}\\
    H[\phi,\tilde{\phi}]&=\int\mathrm{d}\vec{x}\left\{\tilde{\phi}(A+D\nabla^2\phi)+\frac{1}{2}
B
\tilde{\phi}^2\right\}
    \end{align}    
\end{subequations}
The most probable trajectories are the stationary points of
$S$, which are determined by
the Hamiltonian equations:
\begin{align}
    \label{eq:hamilton_eq_1}
    \partial_t\phi
    &=
    \phantom{-}
    \frac{\delta H
    }{\delta \tilde{\phi}
    } = 
   \phantom{-}
    A
    + D\nabla^2\phi + B
    \tilde{\phi} \\
    \label{eq:hamilton_eq_2}
    \partial_t\tilde{\phi}
    &= -\frac{\delta H
    }{\delta \phi
    } = -\tilde{\phi}
    \frac{\delta A}{\delta \phi}
    - D\nabla^2\tilde{\phi} - \frac{1}{2}\tilde{\phi}^2
    \frac{\delta B}{\delta \phi}
\end{align}
The steady-state solution $\tilde{\phi}=0$
of (\ref{eq:hamilton_eq_2}) corresponds to 
deterministic relaxation, while solutions with $\tilde{\phi}\ne 0$ describe 
noise-activated trajectories.

\section{Mean-Field theory}
\begin{figure}
    \centering
\hspace{-0.25cm}\includegraphics[width=0.95\columnwidth]{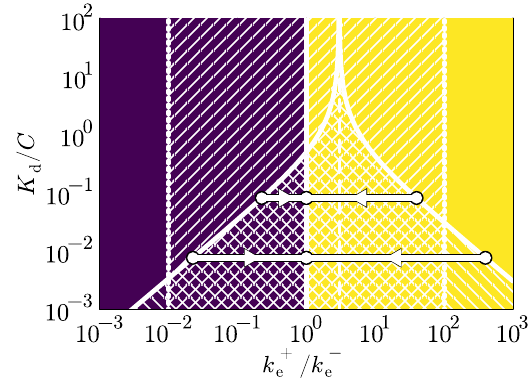}
    \caption{
    Adiabatic relaxation of the catalytic ratio $\rho=k_\mathrm{e}^+/k_\mathrm{e}^-$ in the $(\rho, K_\mathrm{d}/C)$ plane, for $K_\mathrm{m}/C=10^{-2}$, $\rho_0=3$, and $K^\pm_\mathrm{d}=K_\mathrm{d}$. 
    The physically attainable states are restricted to the region between the solid boundary lines [Eq.~(\ref{eq:phys_strip})], which is centered at $\rho=\rho_0$ (dashed vertical line). Points on these boundaries represent uniform states, while the $\rho=1$ solid vertical line corresponds to phase-coexistence steady states. 
    The $+$ and $-$ phases are favored in the yellow ($\rho>1$) and purple ($\rho<1$) regions, respectively. The bistable region is enclosed by dotted lines [Eq.~(\ref{eq:bistability})]. 
    The arrows illustrate the slow adiabatic drift as the system relaxes toward $\rho=1$; the dependence of $\rho$ on the order parameter $\langle\phi\rangle_\mathrm{s}$ is governed by Eq.~(\ref{eq:alphas}). 
    For instance, a uniform state with $\langle\phi\rangle_\mathrm{s}=-c$ corresponds to the right boundary of the physical region (rightmost white dots). Here, random fluctuations trigger the growth of the competing $+$ phase, which recruits corresponding enzymes from the reservoir. This depletion reduces 
    $\rho$, thereby inducing an adiabatic drift toward the coexistence line (leftward arrows). 
    Decay towards the phase-coexistence state occurs via homogeneous nucleation (upper 
    leftward
    arrow) or linear instability (lower 
    leftward
    arrow), depending on whether the initial uniform state lies within or outside the bistable region. An analogous behavior, mirrored across the $\rho=1$ line, is observed for uniform states with $\langle\phi\rangle_\mathrm{s}=+c$ (rightward arrows).
    }
\label{fig:dynamical_pd}
\end{figure}

Setting the response field $\tilde{\phi}=0$
in (\ref{eq:hamilton_eq_1}) and (\ref{eq:hamilton_eq_2})
yields the deterministic dynamics for 
the field $\phi$ in the form of a time-dependent Ginzburg-Landau equation:
\begin{equation}
    \label{eq:TDGL}
\partial_t\phi
(\vec{x},t)
=  
    -\frac{\delta\mathcal{F}
    [\phi]
    }
    {\delta\phi
    (\vec{x},t)
    }
    \end{equation}
    where the effective energy functional
    \begin{eqnarray}
\label{eq:effectiveenergy}
\mathcal{F}&=&\int
\left[
\frac{D}{2}\left|\nabla\phi\right|^2
+V(\phi)
\right]
\mathrm{d}\vec{x} 
\end{eqnarray}
with effective potential
    \begin{eqnarray}
   V(\phi)&=&-\int A(\phi) 
   \mathrm{d}\phi
   \label{eq:potential}
\end{eqnarray}
plays a role analogous to that of the free energy in equilibrium systems. 
Here, the slowly varying spatial average $\langle\phi\rangle_\mathrm{s}$ is treated as a fixed parameter, a condition we will later relax by allowing it to vary adiabatically.
To determine the mean-field phase diagram of the system, we study the behavior of the effective potential $V(\phi)$ as a function of kinetic rates and chemical concentrations.
For simplicity, we set
$K^\pm_\mathrm{m}=K_\mathrm{m}$ and neglect basal conversions 
since we focus on a regime where enzymatic processes dominate ($k_\mathrm{e} \gg k_\mathrm{b}$).
The system exhibits a
{\it bistability region} where the potential
$V(\phi)$ has two minima, corresponding to stable phases
$\phi=\pm c$, and a maximum at the unstable point 
$\phi_0=\left(1+2 \frac{K_\mathrm{m}} 
{C}\right)\frac{k^-_\mathrm{e}-k^+_\mathrm{e}}
{k^+_\mathrm{e}+k^-_\mathrm{e}}\,c$,
where $C=c/\gamma$.
This region (Fig.~\ref{fig:dynamical_pd},  
hatched stripe) 
is defined by the condition:
\begin{equation}
\label{eq:bistability}
    \frac{ K_\mathrm{m}}{ K_\mathrm{m}+C} \leqslant \frac{k^+_\mathrm{e}}{k^-_\mathrm{e}} \leqslant \frac{ K_\mathrm{m}+C}{ K_\mathrm{m}}
\end{equation}
The bistability region widens for smaller values of the Michaelis
constant $K_\mathrm{m}$, i.e., when enzymes operate close to the saturation regime.
For
$k^+_\mathrm{e}>k^-_\mathrm{e}$,
the unstable maximum is located at
$\phi_0<0$, and
the global minimum 
of the potential $V$
is at $\phi=c$, favouring the~$+$~phase (Fig.~\ref{fig:dynamical_pd}, yellow stripe). 
Conversely,
for 
$k^+_\mathrm{e}<k^-_\mathrm{e}$, 
the unstable maximum is at
$\phi_0>0$ and
the global minimum is at $\phi=-c$,  favouring the~$-$~phase  (Fig.~\ref{fig:dynamical_pd}, 
purple stripe).
When 
$k^+_\mathrm{e}=k^-_\mathrm{e}$ the potential $V$ is symmetric, the two phases are equally favored, and 
phase coexistence is possible at the steady state.

\begin{figure*} [t]
    {\centering 
             (a) \hspace{3.5in} 
             (b)\hfill \phantom{.} \\ 
             ~\hspace{0.8cm}
             \includegraphics[width=6.7in]{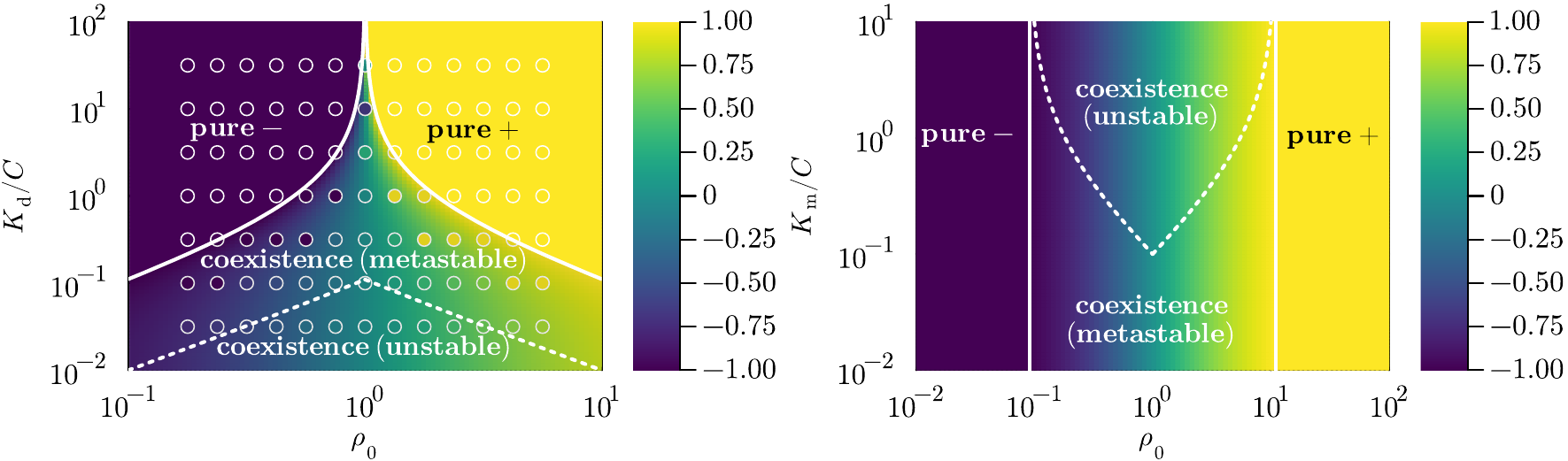}    
             (c) \hfill \phantom{.} \hfill \\
             \vspace{-0.30cm}\hspace{0.1cm}
             \includegraphics[width=6.15in]{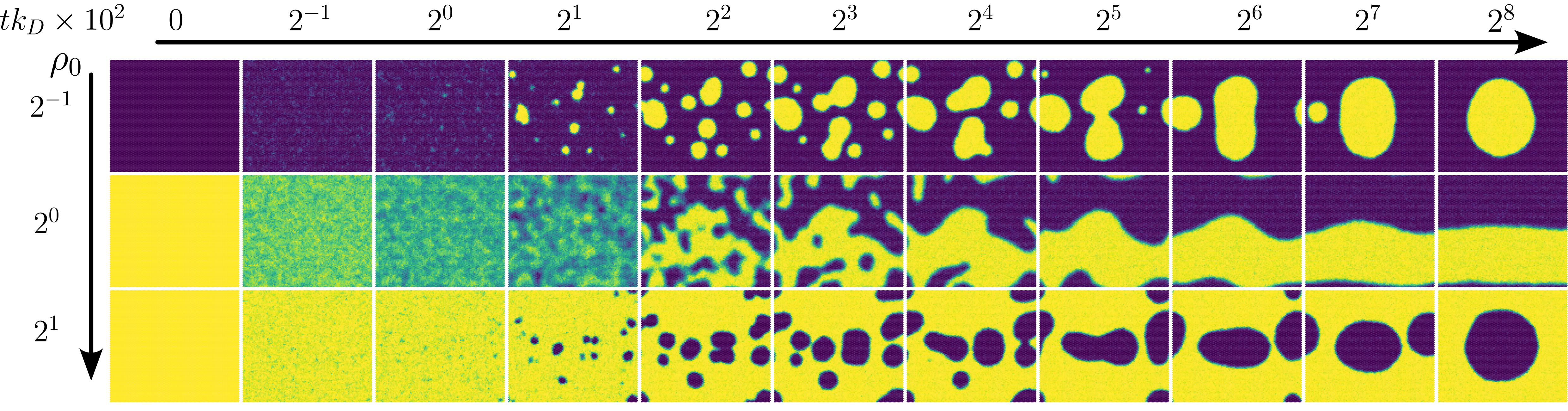}
             
            (d) \hfill \phantom{.} \\
            \vspace{-0.15cm}~\hspace{0.4cm}
            \includegraphics[width=6.12in]{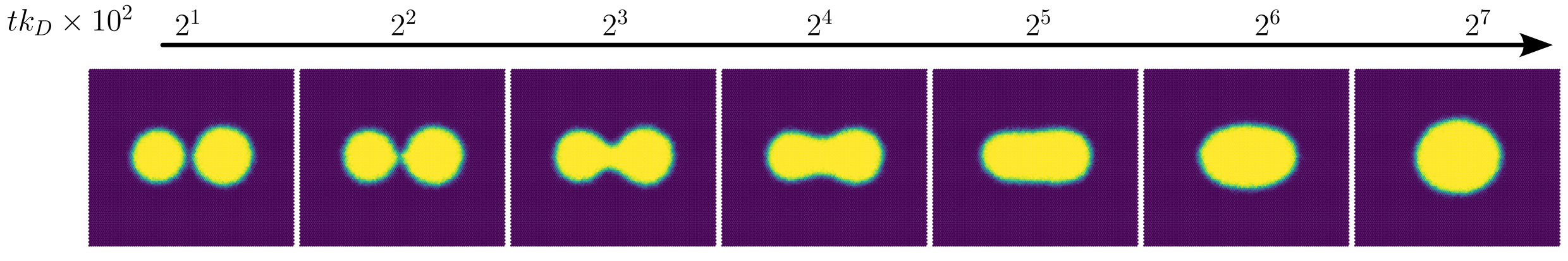}  }
    \caption{{(a,b)} Steady-state phase diagram in the 
    reduced parameter space
    $(\rho_0, K_\mathrm{d}/C,K_\mathrm{m}/C)$.
    Panel (a) shows a projection on the $(\rho_0,K_\mathrm{d}/C)$ plane with
 $K_\mathrm{m}/C$ fixed at~0.1.
   Panel (b) shows a projection on the $(\rho_0,K_\mathrm{m}/C)$ plane with
 $K_\mathrm{d}/C$ fixed at~0.1. 
    We considered here $K^+_\mathrm{d}=K^-_\mathrm{d}=K_\mathrm{d}$ and $K_\mathrm{m}^+=K_\mathrm{m}^-=K_\mathrm{m}$. The solid white lines, defined implicitly by 
    the conditions
    $\rho_0=\rho_\pm$
    [Eqs.~(\ref{eq:def_rhos}),(\ref{eq:phase_coex})]
    from
    the mean-field 
    theory,
    separate 
    regions 
    of pure phases from
    regions of phase coexistence.
    The dashed white lines
    distinguish
    metastable from unstable regions.
    The background color represents the 
    theoretical prediction for
    the steady-state order parameter
    $\langle\phi\rangle_{\mathrm{s},\infty}/c$
    from Eq.~(\ref{eq:phi_eq}). 
    The color inside each circle denotes the  measured $\langle \phi\rangle_\mathrm{s}/c$ in the final time $t=10^5 k_D^{-1}$ of a numerical simulation with the corresponding parameters.
    (c)
    Time evolution of field configurations from 
    simulations with different $\rho_0$ values, 
tuned via the ratio   
$E^+_\mathrm{tot}/E^-_\mathrm{tot}$. 
Here,
$K_\mathrm{d}/C=0.1$.
The mechanism of phase separation depends 
on $\rho_0$,
and can proceed 
either via nucleation or linear instability,
as predicted in panels (a,b). 
The colorscale is the same as in the phase diagram above.
    Simulations were performed with $k_\mathrm{b}\sim10^{-3} k_\mathrm{e}^\mathrm{max}$ (see App.~\ref{app:numerical}).
     (d)
Domain coalescence driven by interface minimization. The simulation shows the merging of two initially separate domains,
driven by minimization of the effective energy
$\mathcal{F}$ concentrated at phase interfaces. }
 \label{fig:pd_new}
\end{figure*}

It is important to recall now that
the $k^\pm_\mathrm{e}$ rates are themselves
slowly varying functions of the field configuration, through its 
spatial average $\langle\phi\rangle_\mathrm{s}$. 
Therefore, not all ratios 
$k^+_\mathrm{e}/k^-_\mathrm{e}$
are physically
realizable, but only those compatible
with the constraint 
$-c\leqslant\phi\leqslant c$. This defines the 
{\it physical region}:
\begin{equation}
    \label{eq:phys_strip}
    \frac{\rho_0}{\rho_+}
        \leqslant
    \frac{k^+_\mathrm{e}}{k^-_\mathrm{e}} 
    \leqslant 
    \frac{\rho_0}{\rho_-}
\end{equation}
(Fig.~\ref{fig:dynamical_pd},
reverse 
hatched region)
where
\begin{eqnarray}
    \label{eq:def_rhos}
    \rho_0=\frac{k^+_\mathrm{c} E^+_\mathrm{tot}}{k^-_\mathrm{c} E^-_\mathrm{tot}}, \  
    \rho_+ =
    \frac{ K^+_\mathrm{d}+C}{ K^-_\mathrm{d}}, \  \rho_- =
    \frac{ K^+_\mathrm{d}}{ K^-_\mathrm{d} +C} 
\end{eqnarray}
A key implication is that the phase-coexistence steady state
$k^+_\mathrm{e}/k^-_\mathrm{e}=1$
is compatible with the physical constraint 
(\ref{eq:phys_strip})
only if the 
catalytic ratio
$\rho_0$
satisfies:
\begin{equation}
    \label{eq:phase_coex}
    \rho_- \leqslant \rho_0 \leqslant \rho_+
\end{equation}
The above conditions 
allow us to 
derive the
steady-state phase diagram
of the system [Figs.~\ref{fig:pd_new}(a) and \ref{fig:pd_new}(b)]. This diagram is defined in the reduced parameter
space 
spanned by three dimensionless combinations of kinetic rates and molecular concentrations: 
$(\rho_0, K_\mathrm{d}/C,K_\mathrm{m}/C)$.
The surfaces
defined by
$\rho_0=\rho_\pm$ [solid lines in Figs.~\ref{fig:pd_new}(a) and \ref{fig:pd_new}(b)],
divide the parameter space in three regions, corresponding to steady states of either the pure $+$~phase, the pure $-$~phase, or a coexistence of both phases.
The width of the phase-coexistence
region is independent of 
the Michaelis
constant
$K_\mathrm{m}$ [Fig.~\ref{fig:pd_new}(b)]
but widens for smaller dissociation constants~$K_\mathrm{d}$ [Fig.~\ref{fig:pd_new}(a)],
i.e. when enzymes from the reservoir have higher affinity for the scaffold.

The phase-coexistence region 
in its turn
is divided in two subregions,
corresponding to 
distinct pathways of
relaxation 
toward the
steady state.
Starting from one of the uniform initial states that are at the extremes of the physical region,
$\langle\phi\rangle_\mathrm{s}=\pm c$,
the system can exhibit
two qualitatively different relaxation mechanisms.
If the initial uniform state is whithin 
the bistability region, 
it is metastable (as illustrated in Fig.~\ref{fig:dynamical_pd}, and detailed in its caption)
and decays via nucleation of droplets of the globally favored phase [Fig.~\ref{fig:pd_new}(c)].
Conversely, if the initial uniform state is outside the bistability region, it is linearly unstable, and phase separation is triggered spontaneously by small fluctuations.
The metastable and unstable regions are separated by the surfaces [Figs.~\ref{fig:pd_new}(a) and \ref{fig:pd_new}(b), dashed lines]:
\begin{equation}
    {\rho_0}
    =\frac{K_\mathrm{m}+C}{K_\mathrm{m}}\rho_-, \qquad
    {\rho_0} 
    =\frac{K_\mathrm{m}}{K_\mathrm{m}+C}\rho_+
\end{equation}

A global negative feedback mechanism governs the long-term dynamics.  
The favored phase is determined by the ratio
$\rho = k_\mathrm{e}^+/k_\mathrm{e}^-$, which, according to Eq.~(\ref{eq:alphas}), 
is a decreasing function of the spatial average  $\langle\phi\rangle_\mathrm{s}$ of the order parameter.
This establishes a self-regulating loop: as the growth of $+$ phase domains 
increases $\langle\phi\rangle_\mathrm{s}$, it simultaneously drives the ratio 
$\rho$ downward. This process induces the adiabatic drift illustrated in 
Fig.~\ref{fig:dynamical_pd} (white leftward arrows), pushing the system 
asymptotically toward 
phase-coexistence, 
where $\rho = 1$ and 
catalytic actions are balanced. Physically, the expansion of a stable phase 
depletes the reservoir of available enzymes of the corresponding type, 
recruiting them from the cytosol and slowing further growth until the system 
relaxes into 
a-phase-coexistence steady state. Here,
the spatial average of the field reaches a value
\begin{align}
    \label{eq:phi_eq}
    \langle \phi \rangle_{\mathrm{s},\infty}
    = 
    \frac{\left(1+ 2 
    \frac{K^-_\mathrm{d}}{C}\right)\rho_0-\left(1+2  \frac{K^+_\mathrm{d}}{C}\right)}{\rho_0+1}\,c
\end{align}
 which determines the fraction of the scaffold occupied by each phase (see App.~\ref{app:phase_diag_details}).
During relaxation to this steady-state asymptotics
the variation of $\langle\phi\rangle_\mathrm{s}$ becomes slower and slower, 
which justifies 
a posteriori 
the
adiabatic assumption.
A similar argument validates
the 
adiabatic assumption
during the nucleation of small domains of the stable phase within a metastable background.

\begin{figure}
    \centering
    \includegraphics[width=3in]{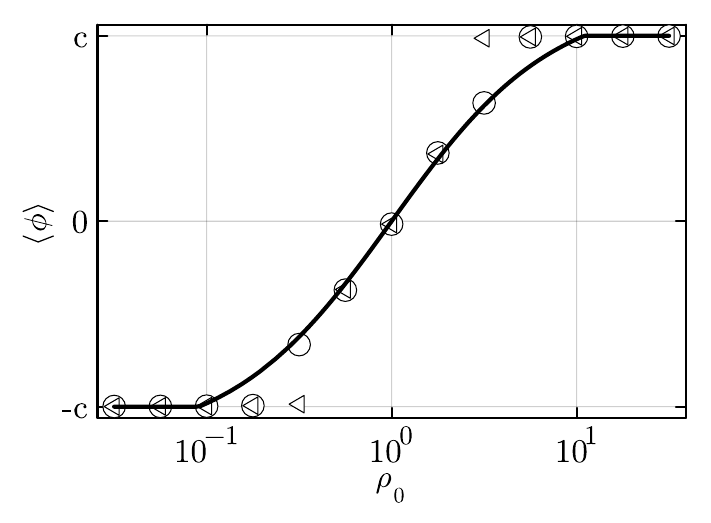}
    \caption{Numerical measurements (symbols) of 
    the steady-state
    average
    $\langle \phi \rangle_\mathrm{s}$ compared with the theoretical prediction (solid line).
    The parameter
    $\rho_0$ is varied
    by independently
    tuning either the catalytic rates $k_\mathrm{c}^\pm$ (circles) or the total enzyme
    concentrations
$E_\mathrm{tot}^\pm$ (diamonds). 
Simulations 
were performed 
with $K_\mathrm{d}/C=0.1$ and $K_\mathrm{m}/C=0.1$.
    }
    \label{fig:phi_av}
\end{figure}

We tested these mean-field predictions with numerical simulations 
of the 
full
stochastic process [Eqs.~(\ref{eq:R1})--(\ref{eq:R3})] on a discrete lattice using the Gillespie algorithm (see 
App.~\ref{app:numerical}
for simulation details and parameters).
Our numerical exploration of the phase diagram shows good qualitative agreement with 
the mean-field prediction 
[Fig.~\ref{fig:pd_new}(a) and Fig.~\ref{fig:phi_av}]. 
However, a quantitative mismatch occurs near the phase boundaries.
This discrepancy arises because the escape time from metastable uniform states diverges near these boundaries (see Sect.~\ref{sec:fluctuations}). 
Since simulations were run for a finite time, 
configurations near the predicted
boundaries 
(corresponding to some of the symbols near $\pm c$ in Fig.~\ref{fig:phi_av})
did not reach the true steady state over the simulation time. 
In contrast, simulations deep within the phase-coexistence region agree very well with the theory. 
The data 
in Fig.~\ref{fig:phi_av},
obtained
by separately varying the 
catalytic 
rate $k_\mathrm{c}^+$ and the
total enzyme 
number $E_\mathrm{tot}^+$, 
confirm that phase coexistence is 
controlled by their product within the dimensionless ratio $\rho_0$ [Eq.~(\ref{eq:def_rhos})].
It is worth noting here that for larger dissociation rates $K_\mathrm{d}$, fewer enzymes are bound to the scaffold, the effective catalytic rates $k_\mathrm{e}^\pm$ decrease, and the time required to reach the steady state
in the simulations
increases accordingly. 

\begin{figure}
    \centering
    \includegraphics[width=8.5cm]{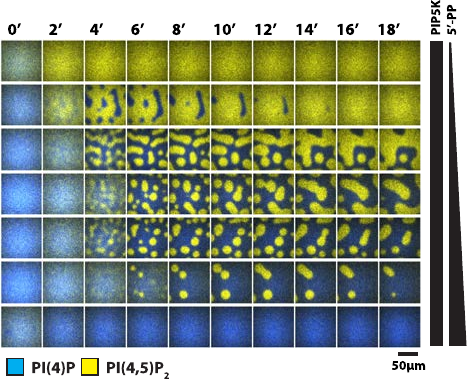}
    \caption{
(Reproduced from Ref.~\cite{HHL+19}, which is published under the CC BY-NC-ND 4.0 license (\href{https://creativecommons.org/licenses/by-nc-nd/4.0/}{https://creativecommons.org/licenses/by-nc-nd/4.0/})).
    Time evolution of lipid domains in an
    \textit{in vitro} reconstituted kinase-phosphatase reactions system. 
    Two antagonistic enzymes, the PIP5K kinase and INPP5E phosphatase, drive the interconversion of the membrane lipids PIP(4,5)P and PI(4)P; these 
    correspond to $E^+$, $E^-$, $\phi^+$ and $\phi^-$ in our model, respectively. Rows from top to bottom correspond to increasing concentrations of INPP5E (5'-PP gradient bar) 
    at a fixed concentration of PIP5K, thereby varying the enzyme concentration ratio.
    Starting from the same initial conditions, the onset of phase separation and a subsequent coarsening process are observed over time for a range of these ratios.
    Consistent with our 
    theoretical prediction [Eq.~(\ref{eq:def_rhos})] and numerical results [see Fig.~\ref{fig:pd_new}(a) and Fig.~\ref{fig:pd_new}(c)], this ratio control the membrane fraction occupied by each phase.
    }
    \label{fig:groves}
\end{figure}

This theoretical scenario qualitatively 
reproduces
experimental
observations 
in a reconstituted kinase-phosphatase system~\cite{HHL+19}.
There,  
a pair of
antagonistic enzymes, a
lipid kinase and
a phosphatase,
catalyze the interconversion between 
the 
lipid states PI(4)P and PI(4,5)P$_2$ on a membrane surface. 
The kinase
exhibits a 
positive feedback by preferentially binding to its product, 
thereby recruiting more kinase to the membrane, 
and the phosphatase is 
genetically 
engineered to implement an analogous feedback mechanism.
Our theoretical phase diagram 
is compatible with 
their
experimental observations,
in particular with the
time-lapse data
obtained by varying enzyme concentrations (see Fig.~\ref{fig:groves}).
Notably, 
the active character
of the phase-ordering process 
is confirmed by their
experimental 
observation
that
domains disappear upon ATP removal
(see Fig.~S8 from Ref.~\cite{HHL+19}).

Furthermore, our theoretical framework is consistent with observations of Rab5 domain formation on synthetic membranes~\cite{CLS+20}. 
In this system, which plays a crucial role in endocytic protein sorting,  the small GTPase Rab5 acts as a two-state molecule, switching between inactive (GDP-bound) and active (GTP-bound) states. The Rabex5/Rabaptin5 complex functions as the catalytic unit that binds the active state (GTP-Rab5) and promotes the transition from the inactive to the active state, thus creating a positive feedback loop. 
 By varying the concentration of the catalytic unit, they observe a transition from a homogeneous state to one with stable domains (see Figs.~3B-D from Ref.~\cite{CLS+20}), 
 which closely mirrors
 the phase transition predicted by our theoretical phase diagram.

The dynamics of sufficiently large phase-separated domains is nearly deterministic and governed by 
Eq.~(\ref{eq:TDGL}),
which
shows 
that the 
system evolves to minimize
the effective energy in Eq.~(\ref{eq:effectiveenergy}).
At large times, this leads to relaxation to the stationary state
\begin{equation}
    0  =  \frac{\delta \mathcal{F}[\phi]}{\delta \phi(\vec{x},t)} =  - D \nabla^2 \phi(\vec{x},t) +
    V' (\phi(\vec{x},t))
\label{eq:statstat}
\end{equation}
For a flat interface 
at coexistence (where the potential is symmetric, $\phi_0=0$,
and $V(\pm c)=0)$ separating regions occupied by the $-c$ and $+c$ phases, this is readily integrated along a coordinate $x$ transverse to the interface
to give
\(
    \frac{D}{2} 
    (\partial_x \phi)^2  =  
    V (\phi)
\).
Approximating $V (\phi)$ with a fourth order polynomial (see App.~\ref{sec:polyn}), this can
be solved explicitly to obtain the interface profile $\phi (x) = c
\tanh (x / w)$, exhibiting the width:
\begin{equation} 
\label{eq:width}
    w  =  \frac{4 Dc^2}{3 \sigma}
\end{equation}
where $\sigma$ is an effective interfacial tension  (effective energy per unit interface
length or area, depending on whether the scaffold is two- or
three-dimensional), expressed in terms of nonequilibrium kinetic
rates:
\begin{eqnarray}
  \sigma & = & \int_{- \infty}^{+ \infty} \left[ \frac{D}{2} (\partial_x
  \phi)^2 + V (\phi) \right] \!\mathd x  \label{eq:inttens}\\
  & = & D \int_{- \infty}^{+ \infty} (\partial_x \phi)^2 \, \mathd x \\
  & = & \int_{- c}^{+ c} \sqrt{2 DV (\phi)} \; \mathd \phi \\
  & = & \frac{4}{3}  \sqrt{D (k_e^+ + k_e^-) g (K_m / C)}\; c^2 
  \label{eq:sigma}
\end{eqnarray}
and:
\begin{eqnarray}
g(\kappa)&=&1+4\kappa(1+\kappa)\ln\frac{4\kappa(1+\kappa)}{(1+2\kappa)^2}
\label{eq:kappa}
\end{eqnarray}
is a  monotonically decreasing function 
with
$g(0)=1$ and
$g(\kappa)\sim\kappa^{-2}$
for large 
$\kappa$. 
Therefore, the interfacial tension is larger 
when enzymes operate near their saturation regime
(smaller $K_\mathrm{m}$).
This decrease of
$\sigma$
with
large~$K_\mathrm{m}$
is a genuine consequence of 
saturation-induced nonlinearity,
as it persists even when the
limit is
taken at constant
$k_\mathrm{e}^\pm/K_\mathrm{m}$ to normalize for the reduced reaction rate~(\ref{eq:alphas}). 

Together, Eqs. (\ref{eq:width}) and (\ref{eq:sigma}) show that 
enzymes working 
near
saturation 
and/or with faster catalytic rates
produce sharper interfaces. 
This can be intuitively understood as follows: self-reinforcing catalytic conversions induce effective interactions between scaffold-bound molecules. More frequent catalytic conversions, driven by either larger $k_\mathrm{c}^\pm$,
smaller $K_\mathrm{m}$, or both, induce stronger effective interactions and result in higher interfacial tension and sharper interfaces.
Importantly, these effective interactions are purely kinetic
and do not require sustained molecular proximity. In this context, the high interfacial tension is a macroscopic manifestation of the rate at which the system dissipates energy to maintain the compositional gradient against diffusive mixing.

\begin{figure}[!ht]
    \centering
    (a) \hfill \phantom{.} \\
\vspace{-0.5cm}\hspace{-0.2cm}\includegraphics[width=3.03in]{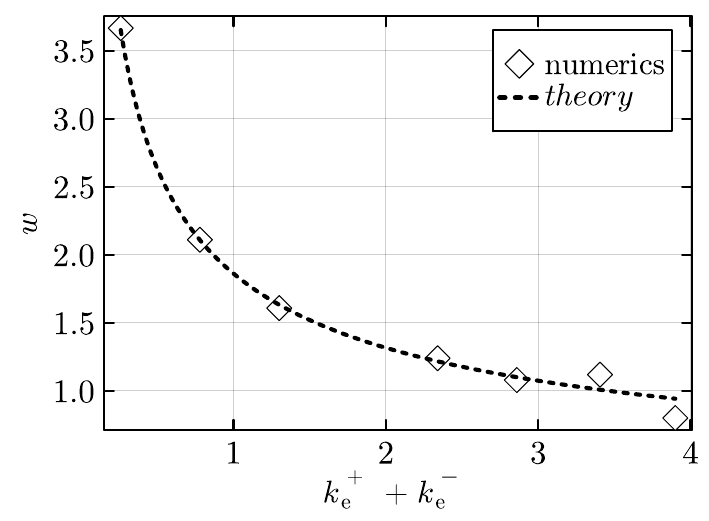}
  \\
  (b) \hfill \phantom{.} \\
\vspace{-0.5cm}\hspace{0.05cm}\includegraphics[width=2.9in]{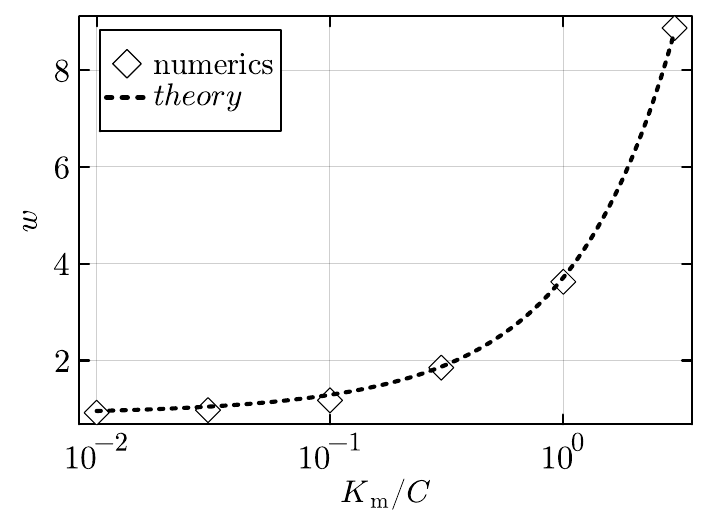}
    \caption{
    Numerically
    determined
    interface width $w$ (in lattice units) for circular domains at the
    steady
    state.
    (a) Interface width 
    as a function  of
    the total catalytic rate
    $k^+_\mathrm{e}+k^-_\mathrm{e}$. 
    (b)~Interface width as a function of the Michaelis
    constant~$K_\mathrm{m}$. Solid curves are fits to Eq.~(\ref{eq:width}),
     with the prefactor 
     as a free parameter. Error bars are 
     smaller than the plot markers.
         The interface width was determined by computing the radial average of the field around a stationary domain and fitting the resulting profile to a hyperbolic tangent.
     Parameters are listed in Table~\ref{table:parvalue} of App.~\ref{app:numerical}.
    } 
\label{fig:interface_width}
\end{figure}

The interfacial width can 
be used to experimentally
measure the interfacial tension via Eq.~(\ref{eq:width}),
provided the other parameters are known.
Our direct numerical measurements of $w$ confirm its predicted dependence on the key catalytic parameters, 
$k_\mathrm{e}^\pm$ and 
$K_\mathrm {m}$
(Fig.~\ref{fig:interface_width}).

Sharp boundaries and high interfacial tension are typically associated with strong, site-specific molecular interactions~\cite{Mus22}.
Our 
active scenario, which is  founded on
intrinsically site-specific enyzmatic interactions,  
demonstrates that
interface sharpness is governed not merely by binding strength,
but also decisively by 
the {frequency} of catalytic events.

Since the effective energy 
$\mathcal{F}$
from 
Eq.~(\ref{eq:effectiveenergy})
is concentrated at phase interfaces,
the dynamics of large 
domains is 
driven
by 
the minimization of the total interface length.
This 
leads to the
coalescence of nearby domains into larger ones to reduce the interface length, as illustrated in simulations where two initially separate domains merge
[Fig.~\ref{fig:pd_new}(d)].

\begin{figure}
    \centering
\includegraphics[width=3.0in]{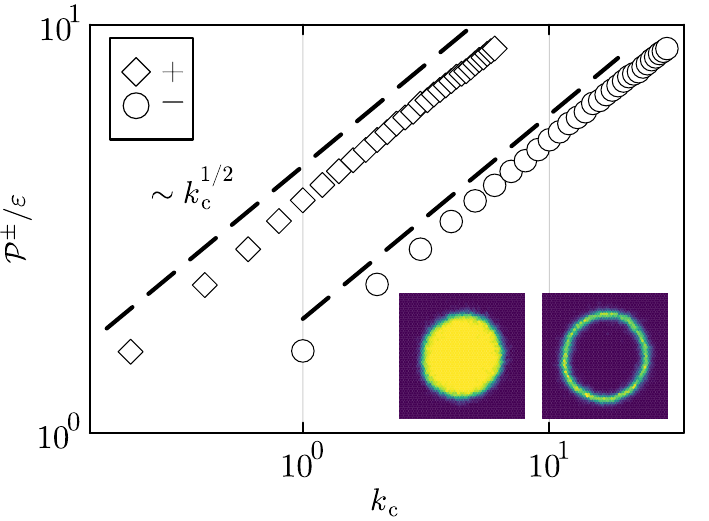}
    \caption{
    Power consumption per unit interface length as a function of the catalytic rate 
    $k_\mathrm{c}$
    for both enzyme species ($+$:~diamonds, $-$:~circles).
    Numerical 
    measurements of the frequency of catalytic events
    are compared
     with the predicted $\sim k_\mathrm{c}^{1/2}$ 
     scaling 
     (dashed lines).
    Left inset:
    A system configuration
    showing 
    a domain of the $+$ phase (yellow) 
    within the $-$ phase (purple).
    Right inset:
    The corresponding
    spatial distribution of  power consumption,
    where yellow
    indicates
    the maximum value and purple indicates zero. 
    Energy dissipation is
    concentrated at the
    phase interface, where futile cycles of antagonist reactions take place.
    }
    \label{fig:power}
\end{figure}

Overall, these results show that 
a gas of reacting and diffusing particles  can exhibit at
the mesoscopic scale a phenomenology 
nearly
indistinguishable from that of classical phase separation.
This includes domain formation via nucleation or linear instability, and coarsening driven by interface minimization
[Fig.~\ref{fig:pd_new}(c) and \ref{fig:pd_new}(d)]. 
However, two key distinguishing features, which could be used for experimental discrimination,
must be emphasized:

First, since the particles behave as a gas, 
particle exchange across boundaries is rapid. 
The 
stability of 
domains 
does not arise
from
reduced particle mobility,
but from their constant interconversion, driven by enzymatic activity at the interfaces [see Fig.~\ref{fig:power}, 
right
inset].

Second, as a direct
consequence, maintaining the domain structure requires a constant energy expenditure. 
The power consumption
per unit interface length
associated with the interfacial activity of each enzyme type
can be computed by integrating the reaction rates (\ref{eq:reactionrates}) over the
previously derived
flat interface profile, obtaining:
\begin{equation}
\mathcal{P}^\pm
=
\frac{1}{2}\, k^{\pm}_\mathrm{e} \mkern2mu \varepsilon_0\, c \, w \ln \left( 1 +
\frac{C}{K_\mathrm{m}} \right)
\label{eq:power}
\end{equation}
where $\varepsilon_0$ is the energy 
expended in a single catalytic event, which
is typically provided by the hydrolysis of a single ATP molecule \cite{AHJ+22}.
Eq.~(\ref{eq:power}) shows that the power required to maintain a phase interface is proportional to the width of the interface, and is thus inversely proportional to the interface tension. 
Combining Eqs.~(\ref{eq:width}) and (\ref{eq:sigma})--
(\ref{eq:power}) shows that
the power consumption
is a slowly decreasing function of $K_\mathrm{m}$, with a logarithmic divergence for $K_\mathrm{m}\ll C$ stemming from the breakdown of the Michaelis-Menten approximation. 
From the same relations 
we find that the power consumption scales with the square root of the catalytic rates $k_\mathrm{c}^\pm$, a scaling confirmed by numerical measurements
(Fig.~\ref{fig:power}).

\section{Fluctuations}
\label{sec:fluctuations}
\begin{figure} 
    \centering
    \vspace{-0.16cm}
\includegraphics[width=3.0in]{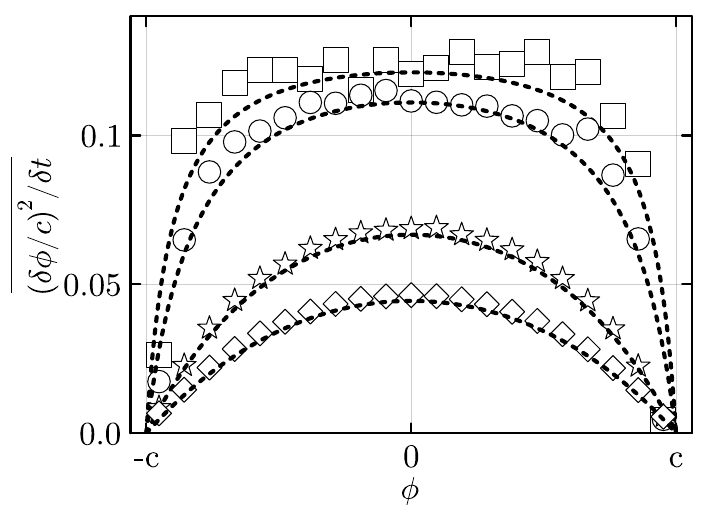}
    \caption{
    Amplitude of field fluctuations as a function of the local field value $\phi$. 
    Symbols show binned numerical data for different Michaelis 
    constants 
    $K_\mathrm{m}/C$: 0.05 (squares), 0.1 (circles), 0.5 (stars),
    and 1 (diamonds).
   Simulations where performed with $k^\pm_\mathrm{b}=0$ 
   and
   a symmetric potential, 
   resulting in 
   a fluctuation
   peak 
   at $\phi=0$. Solid lines 
   correspond to
   the theoretical model [Eq.~(\ref{eq:noiseB})].   
   Error
   bars are smaller than the symbol sizes.
    }
    \label{fig:fluctuations}
\end{figure}

The decay of a uniform state into a 
spatially patterned
state 
is a crucial process 
in biological membranes,
underlying 
phenomena 
such as chemotaxis \cite{RSB+03,KCFR08},
asymmetric cell
division~\cite{CH04,VY18},
proliferation~\cite{HJW+12,WCM+15}, 
immune signaling~\cite{KM06}, and protein sorting~\cite{Stone+17,DKW+21,ZVS+21}.
Linearly unstable uniform states
tend to decay immediately 
and are thus
 unlikely to be directly observable
in real time.
In contrast,
metastable uniform states can 
persist
for long 
periods
before decaying
into a polarized,
phase-separated state,
either spontaneously via homogeneous nucleation 
or 
triggered by
a small external cue~\cite{WL03}.
Fluctuations, 
not considered in 
previous sections, 
are
crucial 
for 
nucleation. 
In our simulations, we observe that 
germs of the stable phase of size smaller
than
a critical size~$R_\mathrm{c}$ 
tend to shrink, 
while 
those larger than~$R_\mathrm{c}$ grow into stable domains,
mirroring 
the phenomenology of  equilibrium phase separation. 
In this nonequilibrium setting, the properties of nucleation 
and the dependence of the critical size on the kinetic parameters of the process can 
be investigated using
large deviation theory~\cite{FW98,T09,Vulpiani+14}.
This approach must 
account
for
the multiplicative nature of 
the
noise [Eqs.~(\ref{eq:noiseB}), (\ref{eq:reactionrates}), (\ref{eq:alphas})], which  itself depends on the field configuration~$\phi$.
Fig.~\ref{fig:fluctuations} compares the analytical prediction
for the 
noise
amplitude~$B(\phi)$ 
[Eq.~(\ref{eq:noiseB})]
with direct simulation measurements. Consistent
 with the 
theoretical limit assumed in the derivation (App.~\ref{app:adiabatic}), the simulations are performed in the regime of negligible enzyme fluctuations (cf. Table~\ref{table:parvalue}), showing
good agreement.

When basal catalysis is switched off ($k_\mathrm{b}=0$), the noise amplitude 
vanishes
in the uniform
states $\phi=\pm c$,
making
them
absorbing: configurations from which the system cannot escape spontaneously.
This is reminiscent of the absorbing consensus states in generalized voter models~\cite{Lig94,NFK96,CMP09,SB09,LDA22}.
In this analogy, 
the $\phi^\pm$ molecules can be seen as voters in two opinion states, and the catalytic enzymes as agents that facilitate opinion changes.

Fig.~\ref{fig:fluctuations} also shows that, according to Eq.~(\ref{eq:noiseB}), noise near uniform states is less suppressed for smaller~$K_\mathrm{m}$, 
(i.e. when enzymes operate near saturation).
Due to the Poissonian character of reaction events, both the amplitude of
field fluctuations and the local energy dissipation are proportional to the overall reaction rate [Eq.~(\ref{eq:noiseB})].
Consequently, $B(\phi)$
is 
minimal
in the bulk of phase-separated domains
($\phi\sim\pm c$) and maximal at the intermediate $\phi$ values characteristic of phase interfaces [Fig.~\ref{fig:fluctuations} and Fig.~\ref{fig:power}, 
right
inset].
\begin{figure}
    \centering
(a) \hfill \phantom{.} \\
\vspace{-0.5cm}\hspace{0.05cm}
\includegraphics[width=3in]{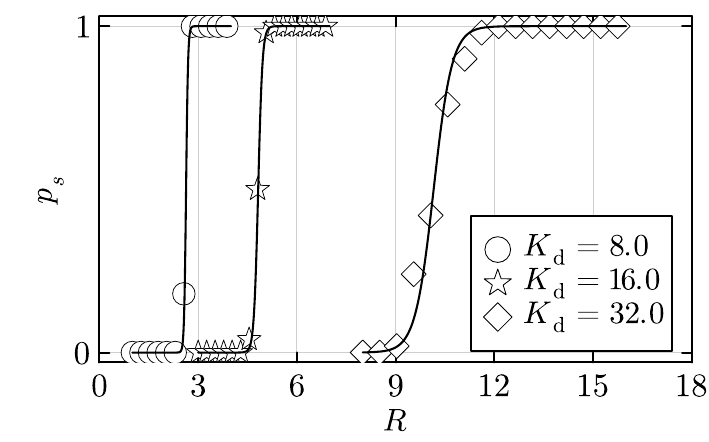}
\\
(b) \hfill \phantom{.} \\
\vspace{-0.5cm}\hspace{-0.1cm}
\includegraphics[width=3.08in]{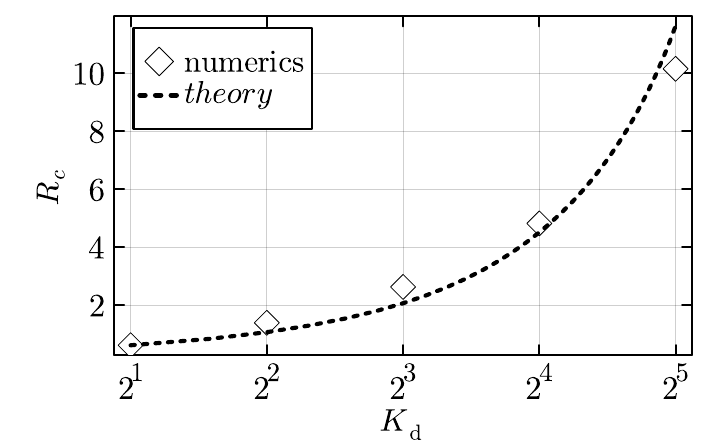}
\\
(c) \hfill \phantom{.}\\
\vspace{-0.5cm}\hspace{0.0cm}
\includegraphics[width=3.1in]{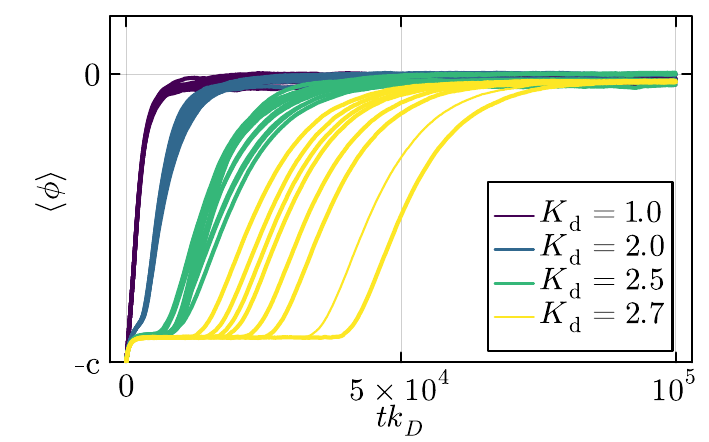}   \caption{
Nucleation dynamics and critical radius.
(a)
Survival probability $p_\mathrm{s}$ 
    of an
    initial domain 
    as a function 
    of 
    its radius~$R$
    (in lattice units)
    for different values of $K_\mathrm{d}$. 
    The critical radius 
    $R_\mathrm{c}$
    for each curve 
    is estimated 
    as 
    the radius 
    where
    $p_s=1/2$. 
    (b)~The critical radius $R_\mathrm{c}$ as a function of the dissociation constant~$K_\mathrm{d}$. The dotted line 
     is
     a fit
     of 
     the theoretical prediction
     from
Eq.~(\ref{eq:critical_radius}), with $k_\mathrm{e}^\pm$ 
from
Eq.~(\ref{eq:alphas}); a~global scaling factor 
     $\approx 3$
     was applied to the theoretical curve to achieve a best fit with the numerical data.
     (c)
     Time evolution of the field average $\langle\phi\rangle_\mathrm{s}$ 
     from simulations
     starting in a homogeneous state.
     For smaller $K_\mathrm{d}$ (purple, blue), 
     the state
     is linearly 
     unstable and 
     separates 
     immediately. For larger $K_\mathrm{d}$ (green, yellow), 
     the state 
     is metastable,
     exhibiting
     a stochastic
     escape time before 
     separation.
     Each color represents multiple
     trajectories.
     All simulations have balanced 
     activity 
     ($\rho_0 = 1$, yielding $\langle\phi\rangle_{\mathrm{s},\infty} = 0$)
     and include a low basal catalysis rate
     $k_\mathrm{b}^\pm=10^{-3}k_\mathrm{e}^\mathrm{max}$ to avoid absorbing states. The initial small ramp 
     is due
     to the 
     slight 
     shift in the minimum of the potential when $k_b^\pm>0$.
     For increasing
     $K_\mathrm{d}$,
     the system lingers in the metastable, homogeneous $-$ state for longer and more variable times until a critical germ 
     of the $+$ phase 
     nucleates and grows deterministically 
     to the steady state.
     }
\label{fig:nucleation_panel}
\end{figure}

The nucleation probability ${P}$ 
for a critical germ is controlled by a large-deviation rate function $S\sim - \log {P}$, 
which depends
on 
the noise amplitude $B(\phi)$. 
The most probable trajectory
out of
a metastable
uniform state is the ``activated" trajectory, i.e. 
the optimal fluctuation
that effectively reverses the deterministic relaxation dynamics, allowing the system to climb the free-energy barrier.
In our case, the trajectory connecting the
uniform state 
to
a circular domain
of radius $R$
is given by the solution of Eqs.~(\ref{eq:hamilton_eq_1}),~(\ref{eq:hamilton_eq_2}) 
evaluated
along the
activated trajectory. Along this path, the auxiliary field satisfies
$\tilde{\phi}=-2
\frac{A+D\nabla^2\phi}{B}$ \cite{Kam23}, yielding:
\begin{equation}
    S[\phi]=\int \mathrm{d}\vec{x}\int\mathrm{d}t\left\{-2 
    \frac{A[\phi]+D\nabla^2\phi}{B[\phi]}\partial_t\phi\right\}
    \label{eq:azione}
\end{equation}
We look for the extremal trajectory in the form of a growing circular droplet,
$\phi(\vec{x},t)=\phi(r-R(t))$, 
allowing
$S$ 
to
be reparametrized as a function of the droplet radius $R$.
For a small critical germ, the dependence of $\langle \phi \rangle_\mathrm{s}$ on $R$ is negligible. We thus treat $\langle \phi \rangle_\mathrm{s}$ as a constant, reducing the functionals $A[\phi], B[\phi]$ to local functions $A(\phi), B(\phi)$ and their functional derivatives to ordinary ones.
In~the 
limit of negligible interface width\
(``thin-wall approximation")
and 
negligible 
basal catalysis,
the maximum
of $S$ 
is found 
at the critical value (cf. App.~\ref{app:nucleation}):
\begin{align}
    R_\mathrm{c}&= -
    \frac{
    \displaystyle
    \int_{-c}^{c}
    \frac{\phi'}{B(\phi)}
    \mathrm{d}\phi
    }
    {
    \displaystyle
     \int_{-c}^{c}
     \left\{
     \frac{2}{D}
     \frac{A(\phi)}{B(\phi)}
     +\frac{\phi'^2
     B'(\phi)}{[B(\phi)]^2}
     \right\}
     \mathrm{d}\phi
    }
\end{align}
Note that Eq.~(\ref{eq:azione}) 
describes 
only the 
noise-driven, 
activated
path
to the 
critical droplet, since 
the 
subsequent
growth
follows a
deterministic path.
Using a polynomial approximation for the potential and noise terms, 
we obtain
an explicit expression for the critical radius:
\begin{align}
\label{eq:critical_radius}
    R_\mathrm{c}&=
    {\frac{3}{20}}
    f({K_\mathrm{m}}/{C})
\frac{\sqrt{D(k^+_\mathrm{e}+k^-_\mathrm{e})}}{|k^+_\mathrm{e}-k^-_\mathrm{e}|}
\end{align}
where
\begin{eqnarray}
\label{eq:fkappa}
  f (\kappa) & = & \frac{1 + 2 \kappa}{\left[ 1 + \frac{12}{5} (\kappa +
  \kappa^2) \right] \sqrt{g (\kappa)}}
\end{eqnarray}
is a
monotonically 
increasing function
with
$f(0)=1$
(see App.~\ref{app:nucleation}).
The critical radius is thus smaller
for 
enzymes operating near  saturation
(smaller $K_\mathrm{m}$)
and for a larger kinetic asymmetry
between the two competing enzymatic
reactions
(larger $|k_\mathrm{e}^+-k_\mathrm{e}^-|$).
It also 
depends
on
kinetic 
rates through 
$k_\mathrm{e}^\pm$.
Eqs.~(\ref{eq:alphas}) and~(\ref{eq:critical_radius}) show that 
$R_\mathrm{c}$
is smaller 
for smaller $K_\mathrm{d}$,
i.e., 
for stronger enzyme affinity for the scaffold.
The nonequilibrium nature of the present result 
is manifested in its dependence on kinetic parameters.
In addition, the peculiar numerical prefactor stems from a distinct contribution proportional
to $B'(\phi)$
that would be absent if noise 
were purely additive.
Finally,
the nucleation radius depends 
on the average composition 
$\langle \phi \rangle_\mathrm{s}$
through 
the effective catalytic rate
$k_\mathrm{e}$, reflecting the constraints imposed by the global conservation law.

To 
numerically
estimate 
$R_\mathrm{c}$
in
the 
full
stochastic process [Eqs.~(\ref{eq:R1})--(\ref{eq:R3})],
we initialized
circular domains of
various
sizes
and 
monitored their evolution.
Smaller domains  shrank and disappeared, 
while larger 
ones
grew to
a stationary state. 
The critical radius was identified as 
the size for which 50\% 
of
domains survived~\cite{FPP+22,ZMF25} 
[see Fig.~\ref{fig:nucleation_panel}(a)].

The analytic expression for the critical radius, Eq.~(\ref{eq:critical_radius}), was derived using several idealizations needed for tractability: the continuum Langevin description, 
circular droplet ansatz,
 thin-wall approximation,  nearly-symmetric effective potential, and the omission of subdominant enzyme noise. Despite these simplifications, direct numerical simulations of nucleation   from a uniform metastable phase confirm the predicted functional dependence of  the critical radius $R_\mathrm{c}$
on the dissociation constant $K_\mathrm{d}$, up to an order-one fitting factor
[Fig.~\ref{fig:nucleation_panel}(b)],
that could also be related to the contribution of nontrivial out-of-equilibrium droplet shape fluctuations~\cite{privatecomm}.

Larger critical sizes
are expected to correspond to lower nucleation rates
for the stable phase within the metastable uniform state.
This is confirmed by simulation results where, with
a low level of basal catalysis ($k_\mathrm{b}^\pm=10^{-3}k_\mathrm{e}^\mathrm{max}$)
maintained to avoid the absorbing state, we find that for
increasing $K_\mathrm{d}$ 
(and thus increasing critical radius), 
the system lingers in the metastable state for progressively longer and more variable times before a critical germ nucleates and grows almost deterministically to the steady state [Fig.~\ref{fig:nucleation_panel}(c)].

\section{Conclusions}
Active phase separation is emerging as a key physical paradigm for intracellular organization, driven by nonequilibrium processes where energy consumption governs the formation of spatial patterns \cite{ZHJ15,WZJ+19,CGA22,DGMF23,BHF20,CN25}.
This framework offers a unifying perspective on biological phenomena, including the assembly of polarized signaling patches on membranes, that have often been described in the past through the lens of ‘‘pattern formation.” Indeed, these processes 
may exhibit the canonical phenomenology of equilibrium phase separation, including domain coalescence, coarsening, and phase ordering driven by interfacial tension, yet they operate through fundamentally different physical mechanisms. 

Here, we have formalized a minimal model of an enzyme-driven phase ordering process, grounded in a well-defined biological module, the counteracting enzymes that govern the activation state of membrane-bound signaling proteins. Our formulation incorporates two realistic ingredients, local reinforcing feedback loops and global constraints imposed by rapid exchange with an enzymatic reservoir, while still remaining analytically tractable.
This structure places the model in the class of locally non-conserved (Model A) dynamics with a global constraint on the total fraction of each molecular state, which is a fundamental mechanism enabling phase coexistence~\cite{GCT+05,GKL+07,OIC+07,HF18,HBF18,BHF20}. 
Mean-field theory yields an explicit phase diagram and closed-form expressions for key observables, including interfacial tension and phase-coexistence boundaries, in terms of basic kinetic rates. The model's simplicity further allows us to move beyond mean-field approximations and analytically investigate fluctuation-driven events. Using large deviation theory, we derived an explicit expression for the critical nucleation radius, which quantifies the stochastic transition from a metastable uniform state to a phase-separated one via the formation of critical droplets.
The quantitative accuracy of our analytical approaches, from mean-field to nucleation theory, was validated through extensive stochastic simulations of the underlying particle-level model, confirming that the simplified theory captures the essential physics of the full system.

In equilibrium phase separation, the onset of phase ordering is controlled by equilibrium parameters like saturation concentrations. Here, an analogous role is played by intrinsically nonequilibrium parameters, such as kinetic rates and enzyme concentrations. The system exhibits abrupt changes in its properties when specific, dimensionless combinations of these parameters exceed a critical threshold, a direct analogue of equilibrium criticality. Thus, the hallmark feature of a robust threshold response, a well-established mechanism for cellular switches between qualitatively different behaviors~\cite{FM98,NP09}, can be preserved and implemented here through intrinsically nonequilibrium means. 

Our analysis yields explicit conditions for phase coexistence, showing it is promoted by a higher enzyme affinity for the membrane and by asymmetries in both enzyme numbers and catalytic activities. Furthermore, we demonstrate that the metastability of uniform states and the sharpness of interdomain interfaces are enhanced when enzymes operate near saturation. Critically, phase ordering is driven here by energy-consuming, nonequilibrium processes. Unlike equilibrium phase separation, the establishment and maintenance of interfaces require continuous energy expenditure. This dissipation is minimal within the bulk of the domains and maximal at the interfaces, where incessant enzymatic ‘‘futile'' cycles sustain the phase separation. Our theory provides closed-form expressions for directly observable quantities, such as interface width, domain area fraction, and critical nucleation radii, offering a set of testable predictions that are directly accessible to targeted experiments.

While the system described here shares much of the phenomenology of equilibrium phase separation, including an effective dynamics that minimizes an effective interfacial energy, its underlying physics is distinct. The interactions driving the ordering process do not require continuous direct 
contact, but are mediated by pointlike, intermittent enzymatic events. This permits the formation of ordered, persistent structures even from a dilute, gas-like phase. In such a scenario, rapid exchange of material between domains and the surrounding environment is expected to be the rule. This rapid exchange, frequently observed in FRAP experiments~\cite{BLH+17,FCWZ19,HHL+19,CLS+20}, and often attributed to the liquid character of condensates, would be even more pronounced for gas-like domains, such as the persistent membrane-associated domains observed in chemotaxing cells \cite{MSU13}.

A central challenge in applying equilibrium phase separation to biomolecular condensates lies in its frequent reliance on nonspecific molecular interactions, which can seem at odds with the highly regulated, combinatorial logic of cellular biochemistry \cite{Mus22}. 
Enzyme-driven phase separation, being intrinsically based on site-specific enzymatic activity, is a universal mechanism that 
intrinsically supports the regulated, combinatorial character inherent to cell physiology. While we have considered here the simplest module consisting of a couple of counteracting enzymes and their membrane-bound substrates, it is quite easy to imagine numerous such modules, involving diverse enzymatic feedback loops, operating in parallel and interacting in various ways 
(a few illustrative examples are given in App.~\ref{sec:multispecies}). 

Qualitatively, the 
presented scenario of
active phase-separation
aligns
with experimental observations on reconstituted membranes. The agreement spans distinct biochemical systems, 
capturing
the evolution of lipid domains governed by a kinase-phosphatase pair~{\cite{HHL+19}} 
as well as
the formation of protein domains driven by a GTPase and its GEF/effector complex~\cite{CLS+20}. The experimentally demonstrated dissolution of domains upon ATP removal~{\cite{HHL+19}} underscores 
in those systems
the essential role of energy consumption, which is a cornerstone of our theoretical framework. The ability of our minimal model to describe such biochemically disparate systems suggests that
its underlying principles,
grounded in nonequilibrium thermodynamics and stochastic kinetics,
constitute
a general physical mechanism for
domain formation on 
molecular scaffolds, 
including cell membranes
as a primary example.

\acknowledgments
We gratefully acknowledge fruitful discussions with Guido Serini, Carlo Campa,
and Andrea Musacchio. Numerical calculations were performed using resources of the Leonardo cluster at CINECA under a CINECA-INFN agreement.
\appendix

\section{Microscopic model}
\label{app:microscopic}
\begin{figure}
    \centering
    \includegraphics[width=\linewidth]{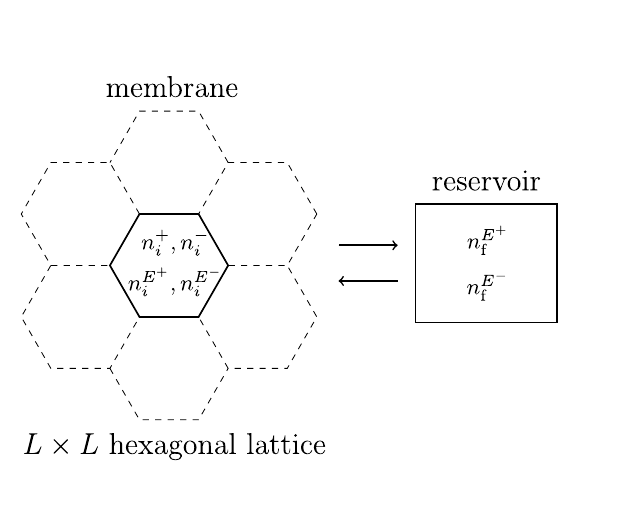}
    \caption{Schematic 
    of the discrete system 
    implemented in the simulations. The scaffold is modeled as an $L \times L$ hexagonal lattice; within each cell $i$,  
    $n_i^\pm$
    and
    $n_i^{E^\pm}$
    denote the number of $\phi^\pm$~molecules and bound 
    $E^\pm$~enzymes, respectively. The 
    unstructured 
    reservoir 
    contains $n_\mathrm{f}^{E^\pm}$~free enzymes.
    }
    \label{fig:lattice_schematic}
\end{figure}

The system, defined by the chemical 
reactions in Eqs.~(\ref{eq:R1})--(\ref{eq:R3}),
together with molecular diffusion, is discretized on a 
$L \times L$ hexagonal lattice with periodic boundary conditions, representing a lipid membrane scaffold exchanging particles with an unstructured reservoir.
The system state at any instant of time
is given by 
the
number of molecules $n^+_i, n^-_i, n_i^{E^+}, n_i^{E^-}$ 
at each site
$i=1,..., L^2$ and
the
number 
$n^{E^+}_{\mathrm{f}},n^{E^-}_{\mathrm{f}}$
of free enzymes in the reservoir 
(see Fig.~\ref{fig:lattice_schematic}).
The following independent Poisson processes are considered:
\begin{enumerate}[label=\alph*)]
    \item Diffusion of a molecule to a neighboring cell with rate $k_D n_i^\pm$.
    \item Diffusion of an enzyme to a neighboring cell with rate $k_D n_i^{E^\pm}$.
    \item 
    Enzymatic
    catalysis 
    converting
    one $\mp$ molecule to one $\pm$ molecule,
    mediated by
    enzymes $E^\pm$ 
    at a site, with rate
    $\frac{{k}^\pm_\mathrm{c} n^{E^\pm}_i \!n^\mp_i}{\gamma\Omega K_\mathrm{m}+n^\mp_i}$, where $K_\mathrm{m}$ is a 
    Michaelis 
    constant.
    \item Conversion of a $\mp$ molecule  to a $\pm$ molecule  with rate $k^\pm_\mathrm{b}n_\mp^i$.
    \item Attachment of 
a free enzyme 
$E^\pm_{\mathrm{f}}$
from the reservoir to a  site $i$, promoted by the presence of its product molecule ($\pm$),
with rate ${k}^\pm_\mathrm{a} n^{E^\pm}_{\mathrm{f}}\!\!n_i^\pm$.
    \item 
    Detachment of a bound $E^\pm$ enzyme from site $i$,
    with rate 
    ${k}^\pm_\mathrm{d} n^{E^\pm}_i\!$.
\end{enumerate}

\section{Langevin dynamics} 
\label{app:micromeso}
In this Appendix we
derive the Langevin equation for the
total concentration
field $c$
by coarse-graining the stochastic process~[Eqs.~(\ref{eq:R1})--(\ref{eq:R3})].
This provides a pedagogical illustration of the methods used and recovers a 
standard result for diffusion processes~\cite{Gar85}.
Consider
the total number of substrate molecules
($+$ species and $-$ species) 
at site $i$,
defined as $N_i=
n_i^++n_i^-$. Let $k_D$ be the microscopic hopping rate 
to any of the $q$ neighboring sites on the lattice,
where each $d$-dimensional site has volume $\Omega$, and the center-to-center distance 
between 
neighboring sites is $h$.
Within an infinitesimal time interval 
$\delta t$, 
we
consider 
independent
single events  
and 
calculate 
the
probabilities 
for 
fluctuations 
to first order in~$\delta t$:

\begin{align*}
    \mathbb{P}[\delta N_i=+1]&=\frac{k_D\delta t}{q}\sum_{j \in \mathcal{N}_i} N_j\\
    \mathbb{P}[\delta N_i=-1]&=k_D\delta t N_i\\
    \mathbb{P}[\delta N_i=+1,\delta N_l=-1]&=\frac{k_D\delta t}{q}\sum_{j \in \mathcal{N}_i} \delta_{j,l} N_l\\
    \mathbb{P}[\delta N_i=-1,\delta N_l=+1]&=\frac{k_D\delta t}{q}\sum_{j \in \mathcal{N}_i} \delta_{j,l} N_i
\end{align*}
where $\mathcal{N}_i$ denotes the set of nearest neighbors of a site $i$ on the lattice, and
the final two expressions account for correlations between changes in neighboring sites due to molecular jumps. From these, we calculate the first two moments of the fluctuations:
\begin{widetext}
    \begin{align}
    \mathbb{E}[\delta N_i]&=\sum_v v\, \mathbb{P}[\delta N_i=v]=\frac{k_D\delta t }{q}\sum_{j \in \mathcal{N}_i} (N_j-N_i)\\
    \mathbb{E}[\delta N_i \delta N_l]&=\sum_{u,v} u\, v \, \mathbb{P}[\delta N_i=u,\delta N_l=v]
    =\frac{k_D\delta t}{q}\left[\delta_{i,l}\sum_{j \in \mathcal{N}_i} (N_j+N_i)-\sum_{j\in\mathcal{N}_i}\delta_{j,l} (N_j+N_i)\right]
\end{align}
To proceed to the continuum limit, we define the concentration field $N_i(t)/\Omega \to c(\vec{x},t)$. 
Let
$\hat{u}_k$ (for $k=1,...,q$) be unit vectors pointing toward neighboring site centers. Lattice symmetry implies $\sum_k \hat{u}_{k}=0$ and 
$\sum_{k}\hat{u}_{k,i} 
\hat{u}_{k,j} 
=q\,\delta_{i,j}/d$. 
Taking the continuum limit,
where
$\delta_{ij}/\Omega\to\delta(\vec{x}_i-\vec{x}_j)$ and $\frac{ 1}{q}\sum_{k=1}^q [c(\vec{x}+h \hat{u}_{k},t)-c(\vec{x},t)]
\approx 
\frac{h^2}{2 d}\nabla^2 c(\vec{x},t)$, 
we 
define 
the 
macroscopic diffusion coefficient $D=\lim_{h \to 0} \frac{k_D h^2}{2 d}$.
The deterministic term 
becomes:
\begin{align}
    \mathbb{E}[\delta c(\vec{x},t)]&=\lim_{\Omega\to0}\frac{\mathbb{E}[\delta N_i]}{\Omega}
    =\frac{k_D\delta t}{q}\sum_{k}^q [c(\vec{x}+h \hat{u}_{k},t)-c(\vec{x},t)] 
    =D\delta t \nabla^2 c(\vec{x},t)
\end{align}
The fluctuation term is derived analogously, yielding:
    \begin{align}
        \mathbb{E}[\delta c(\vec{x},t)\delta c(\vec{x}',t)]&=\lim_{\Omega\to0}\frac{\mathbb{E}[\delta N_i \delta N_l]}{\Omega^2} 
        = 2  D\delta t\,\nabla \cdot \nabla'  [c(\vec{x},t)\delta(\vec{x}'-\vec{x})]
    \end{align}
    \end{widetext}
This recovers the
standard result~\cite{Gar85}.
The corresponding Langevin equation for the total concentration field $c$ is therefore:
\begin{equation}
    \label{eq:langevin_c}
    \partial_t c(\vec{x},t)=D\nabla^2c+\nabla \cdot \vec{\zeta}_D
\end{equation}
where $\vec{\zeta}_D$ is a zero-mean diffusion noise with correlation  
$$\langle \zeta_{D}^i(\vec{x},t)\zeta_{D}^j(\vec{x}',t')\rangle=2Dc(\vec{x},t)\delta_{i,j}\delta(\vec{x}-\vec{x}')\delta(t-t')$$ 
Due to
the 
Markovian nature of the 
underlying microscopic process,
Eq.~(\ref{eq:langevin_c})
must be 
interpreted 
in the It\^{o} sense~\cite{Gar85}.
We have thus obtained a self-consistent, purely diffusional dynamical equation for $c$. 
Note that the continuum limit should be regarded as a convenient notation. It implies a coarse-grained model where the size of a lattice site appears infinitesimal on the macroscopic scale, yet remains large enough to contain a sufficient number of molecules for a statistical description~\cite{Gar85}.

We now apply the same procedure to derive the Langevin equation for the order parameter field,
$\phi$,
starting from the corresponding discrete field
$n_i=n^+_i-n^-_i$. 
Let $n_i^{E^\pm}$ be the number of $\pm$ enzymes at site $i$.
The first two moments of $\delta n_i$ are:
\begin{widetext}
    \begin{align*}
    \mathbb{E}[\delta n_i]&=\left[ 
    \frac{2k^+_\mathrm{c}n^{E^+}_in^-_i}{\gamma\Omega K^+_\mathrm{m}+n^-_i} +
    2
    k^+_\mathrm{b}
    n^-_i 
    -\frac{2k^-_\mathrm{c}n_i^{E^-}n_i^+}{\gamma\Omega K^-_\mathrm{m}+n_i^+}
    -2k^-_\mathrm{b}
    n_i^+
    +\frac{k_D}{q}\sum_j(n_j-n_i)
    \right]\delta t\\
    \mathbb{E}[\delta n_i\delta n_l]&=\left[
    \frac{4k^+_\mathrm{c}n_i^{E^+}n^-_i}{\gamma\Omega 
    K^+_\mathrm{m}+n^-_i}
    +   4k^+_\mathrm{b}
    n^-_i
    -\frac{4k^-_\mathrm{c}n_i^{E^-}n_i^+}{\gamma\Omega K^-_\mathrm{m}+n_i^+}
        -4k^-_\mathrm{b}
        n_i^+
    +\frac{k_D}{q}\sum_j(N_j+N_i)
    \right]\delta_{i,l}\, \delta t 
    -\frac{k_D\delta t }{q}\sum_{j\in \mathcal{N}_i}\delta_{j,l}(N_j+N_i)
\end{align*}
We define the continuum concentration fields:
$   n_i(t)/\Omega\to \phi(\vec{x},t), \ 
    n_i^\pm(t)/\Omega\to \phi^\pm(\vec{x},t),  
    n^{E^\pm}_i(t)/\Omega\to E^\pm(\vec{x},t)
$.
The first two moments for the fluctuations of the field
$\phi$
are then calculated. Taking the continuum limit, they read:
\begin{align}
    \mathbb{E}[\delta\phi(\vec{x},t)]
    &=\lim_{\Omega \to 0}\frac{\mathbb{E}[\delta n_i]}{\Omega}
    \quad\,\,\, =\left( 
    \frac{2k^+_\mathrm{c}E^+\phi^-}{\gamma K_\mathrm{m}^++\phi^-}
    +    2k^+_\mathrm{b}
    \phi^-
    -\frac{2k^-_\mathrm{c}E^-\phi^+}{\gamma K_\mathrm{m}^-+\phi^+}
        -2k^-_\mathrm{b}
        \phi^+
    +D\nabla^2\phi
    \right)\delta t\\
    \mathbb{E}[\delta\phi(\vec{x},t)\delta\phi(\vec{x}',t)]&=\lim_{\Omega \to 0}\frac{\mathbb{E}[\delta n_i\delta n_l]}{\Omega}=\left( 
    \frac{4k^+_\mathrm{c}E^+\phi^-}{\gamma K_\mathrm{m}^++\phi^-}
    +4k^+_\mathrm{b}
    \phi^-
    +\frac{4k^-_\mathrm{c}E^-\phi^+}{\gamma K_\mathrm{m}^-+\phi^+}
     +4k^-_\mathrm{b}
     \phi^+   
    \right)
    \delta(\vec{x}-\vec{x}')\,\delta t\nonumber\\
    &\phantom{=\lim_{\Omega \to 0}\frac{\mathbb{E}[\delta n_i\delta n_l]}{\Omega}==}+2 \delta t D\nabla \cdot \nabla' [c(\vec{x},t)\delta(\vec{x}'-\vec{x})]
\end{align}    
Given these moments, 
and using $\phi^\pm=(c\pm\phi)/2$ to reexpress the 
results
in terms of $c$ and $\phi$, 
the 
Langevin equation 
for the order parameter field 
is~\cite{Gar85}:
\begin{equation}
    \label{eq:langevin_phi_raw}
     \partial_t\phi(\vec{x},t)=D \nabla^2 \phi +A+\sqrt{B} \ \xi_R +\nabla \cdot \vec{\xi}_D
\end{equation}
where 
\begin{align*}
    A&=
    \frac{2k^+_\mathrm{c}E^+(c-\phi)}{2\gamma K_\mathrm{m}^++c-\phi}
    +    k^+_\mathrm{b}
    (c-\phi)
    -\frac{2k^-_\mathrm{c}E^-(c+\phi)}{2\gamma K_\mathrm{m}^-+c+\phi}
        -k^-_\mathrm{b}
        (c+\phi)
    \\
    B&=
    \frac{4k^+_\mathrm{c}E^+(c-\phi)}{2\gamma K_\mathrm{m}^++c-\phi}
    +    2k^+_\mathrm{b}
    (c-\phi)+\frac{4k^-_\mathrm{c}E^-(c+\phi)}{2\gamma K_\mathrm{m}^-+c+\phi}
        +2
        k^-_\mathrm{b}
        (c+\phi)  
\end{align*}

Here,
$\xi_R$ and $\vec{\xi}_D$ are
the 
zero-mean
reactions and diffusion
noise terms,
respectively, with
correlations 
\begin{align*}
    \langle \xi_R(\vec{x},t)\xi_R(\vec{x}',t')\rangle&=\delta(\vec{x}-\vec{x}')\delta(t-t')\\
    \langle \xi_D^i(\vec{x},t)\xi_D^j(\vec{x}',t')\rangle&=2Dc(\vec{x},t)\delta_{i,j}\delta(\vec{x}-\vec{x}')\delta(t-t')
\end{align*}

By the same procedure we derive the Langevin equation for the enzymatic component

\begin{align}
        \label{eq:langevinE}
        \partial_t E^\pm&=D\nabla^2E^\pm+\frac{k^\pm_\mathrm{a}E^\pm_\mathrm{f}(c\pm\phi)}{2}-k^\pm_\mathrm{d}E^\pm+\sqrt{\frac{k^\pm_\mathrm{a}E^\pm_\mathrm{f}(c\pm\phi)}{2}+k^\pm_\mathrm{d}E^\pm} \ \eta_{E^\pm}+\nabla \cdot \vec{\eta}_{D^\pm}
    \end{align}
    
\end{widetext}
where $\eta_{E^\pm}$ and $\vec{\eta}_{D^\pm}$ are
the 
zero-mean
reactions and diffusion
noise terms,
respectively, with
correlations 
    \begin{align*}
        \langle \eta_{E^\pm}(\vec{x},t)\eta_{E^\pm}(\vec{x}',t')\rangle&=\delta(\vec{x}-\vec{x}')\delta(t-t')\\
        \langle\eta_{D^\pm}^i(\vec{x},t)\eta_{D^\pm}^j(\vec{x}',t')\rangle&=2DE^\pm(\vec{x},t)\delta_{i,j}\delta(\vec{x}-\vec{x}')\delta(t-t')
    \end{align*}
    and for the reservoir concentration 
    the following
    conservation law
    holds:
    \begin{equation}
         \partial_t E_\mathrm{f}^\pm=-\frac{1}{\gamma}\langle\partial_t E^\pm\rangle_\mathrm{s}
    \end{equation}
    where $\langle\cdots\rangle_\mathrm{s}$ denotes the spatial average over the scaffold.

    The Langevin equations for the fields $c, \phi,$ and $E^\pm$ [Eqs.~(\ref{eq:langevin_c}),(\ref{eq:langevin_phi_raw}),(\ref{eq:langevinE})] include diffusive noise terms.
    Since
    diffusive noise  scales subdominantly with the coarse-graining length compared to non-conserved noise~\cite{W16},
    these terms 
    will be
    neglected 
    hereafter.
    Moreover,
    the total concentration $c$ is unaffected by interconversion reactions and evolves through a purely diffusive process [Eq.~(\ref{eq:langevin_c})].
    After a relaxation transient, its average reaches a homogeneous value determined by the initial conditions.
    By neglecting the diffusive fluctuations around this mean, we can treat this field as a constant, $c(\vec{x}, t) = c$.
    Under this approximation, the dynamics of the substrate molecules on the scaffold is fully captured by the evolution of the order parameter field, $\phi(\vec{x}, t)$.
    
\section{Adiabatic elimination of fast variables}
\label{app:adiabatic}

    To simplify this set of coupled Langevin equations for the concentration difference $\phi$ and enzymes concentrations $E^\pm$ we 
    perform an adiabatic elimination of fast variables. This allows us to enslave the enzyme degrees of freedom to the molecular concentrations. Specifically, we consider the regime where
    \begin{equation}
        k_D,k_\mathrm{c}\ll k_\mathrm{d}
    \end{equation}
    \begin{widetext}\noindent
    Assuming for simplicity 
    $k_\mathrm{d}^\pm=k_\mathrm{d},k_\mathrm{c}^\pm=k_\mathrm{c}$, our
    system is described by three coupled Langevin equations:
    \begin{equation}
        \begin{cases}
            \partial_{t}E^+	=	k_{\mathrm{d}}\left(\frac{D}{k_\mathrm{d}}\nabla^2E^++\bar{E}^+-E^+\right)+\sqrt{k_{\mathrm{d}}\left(\bar{E}^++E^+\right)}\eta_{E^+}\\
            \partial_{t}E^-	=	k_{\mathrm{d}}\left(\frac{D}{k_\mathrm{d}}\nabla^2E^-+\bar{E}^--E^-\right)+\sqrt{k_{\mathrm{d}}\left(\bar{E}^-+E^-\right)}\eta_{E^-}\\
            \partial_{t}\phi	=	k_{\mathrel{c}}f(\phi,E^\pm)+\sqrt{k_{\mathrm{c}} \ g(\phi,E^\pm)}\eta_{\phi}

        \end{cases}
    \end{equation}
    where $\bar{E}^\pm=\frac{k_\mathrm{a}^\pm E_\mathrm{f}^\pm (c\pm\phi)}{2k_\mathrm{d}^\pm}$ is the stationary value for the deterministic part of the equation for the enzymes,
    and
    $f$ and $g$ are defined by:
    \begin{align*}
        f&=\frac{D\nabla^2\phi+A}{k_\mathrm{c}}
        =\frac{D}{k_\mathrm{c}}\nabla^2\phi+\frac{2E^+(c-\phi)}{2\gamma K_\mathrm{m}^++c-\phi}+\frac{k_\mathrm{b}^+}{k_\mathrm{c}}(c-\phi)
         -\frac{2E^-(c+\phi)}{2\gamma K_\mathrm{m}^-+c+\phi}-\frac{k_\mathrm{b}^-}{k_\mathrm{c}}(c+\phi)\\
        g&=\frac{B}{k_\mathrm{c}}\nonumber 
         =\frac{4E^+(c-\phi)}{2\gamma K_\mathrm{m}^++c-\phi}+\frac{2k_\mathrm{b}^+}{k_\mathrm{c}}(c-\phi)+\frac{4E^-(c+\phi)}{2\gamma K_\mathrm{m}^-+c+\phi}+\frac{2k_\mathrm{b}^-}{k_\mathrm{c}}(c+\phi)
    \end{align*}
    \end{widetext}
    where these terms are derived using the drift and variance from Eq.~(\ref{eq:langevin_phi_raw}) and the dependence on the rate $k_\mathrm{c}$ characterizing the timescale is made explicit.
    The noise has zero mean and correlation
    \begin{align}
        \langle\eta_{\alpha}(t,\vec{x})\eta_{\beta}(t',\vec{x}')\rangle	=	\delta_{\alpha,\beta}\delta(t-t')\delta(\vec{x}-\vec{x}')
    \end{align}
    The diffusion terms in the equations for the enzymes can be neglected over scales $\lambda\gg\sqrt{\frac{k_D\Omega}{k_\mathrm{d}}}$, which in the
    regime $k_D\ll k_\mathrm{d}$ is below the resolution of the lattice.
    Moreover, when $k_\mathrm{d}\gg k_\mathrm{c}$, the field $\phi$ (and consequently $\bar{E}^\pm$) is slowly varying and we can approximate is as a constant on the characteristic timescale of the dynamics of $E^{\pm}$.  
   Introducing the dimensionless time $s=k_\mathrm{c}t$ and the dimensionless parameter $\epsilon=k_\mathrm{c}/k_\mathrm{d}$, we can rewrite the equations as
   \begin{equation}
        \begin{cases}
            \partial_{s}E^+=\epsilon^{-1}(\bar{E}^+-E^+)+\sqrt{\epsilon^{-1}\left(\bar{E}^++E^+\right)}\tilde{\eta}_{E^+}\\
            \partial_{s}E^-=\epsilon^{-1}(\bar{E}^--E^-)+\sqrt{\epsilon^{-1}\left(\bar{E}^-+E^-\right)}\tilde{\eta}_{E^-}\\
            \partial_{s}\phi	= f(\phi,E^\pm)+\sqrt{g(\phi,E^\pm)}\tilde{\eta}_{\phi}

        \end{cases}
    \end{equation}
    where we have rescaled the noise $\tilde{\eta}_\alpha=\eta_\alpha/\sqrt{k_\mathrm{c}}$ so that
    \begin{align}
        \langle\tilde{\eta}_{\alpha}(t,\vec{x})\tilde{\eta}_{\beta}(t',\vec{x}')\rangle	=	\delta_{\alpha,\beta}\delta(t-t')\delta(\vec{x}-\vec{x}')
    \end{align}
    The dynamics of the enzymes is the faster one, therefore $E^\pm$ rapidly relax to the stationary value $\bar{E}^\pm$.
    The stochastic process governing $E^\pm$ is a shifted Feller process, whose stationary distribution is a Gamma distribution~\cite{Fel51,Gar85,CIR85}. 
    We  can quantify the fluctuations $\delta E^\pm$ around the stationary value by expanding 
    $E^\pm(s,\vec{x})=\bar{E}^\pm(\vec{x})+\delta E^\pm(s,\vec{x})$,
    When the local number of enzymes is large, fluctuations are small ($\delta E^\pm\ll\bar{E}^\pm$) and we can linearize the noise term by approximating 
     $\sqrt{\bar{E}^\pm+\delta E^\pm}\approx \sqrt{\bar{E}^\pm}$. 
     This way the process is reduced to a Gaussian Ornstein-Uhlenbeck process
    \begin{equation}
        \partial_s \delta E^\pm=-\frac{1}{\epsilon}\delta E^\pm+\sqrt{\frac{2\bar{E}^\pm}{\epsilon}}\tilde{\eta}_{E^\pm}
    \end{equation}
    with zero mean and correlation
    \begin{equation}
        \langle\delta E^\pm(s,\vec{x})\delta E^\pm(s',\vec{x}')\rangle=\bar{E}^\pm\delta(\vec{x}-\vec{x}')e^{-|s-s'|/\epsilon}
    \end{equation}
    In the limit of small autocorrelation time $\epsilon$, we can approximate the correlation by a delta function
    \begin{equation}
        \langle\delta E^\pm(s,\vec{x})\delta E^\pm(s',\vec{x}')\rangle\xrightarrow{\epsilon\to 0}2\epsilon\bar{E}^\pm\delta(\vec{x}-\vec{x}')\delta(s-s')
    \end{equation}
    To study the effect of these small fluctuations on the slow dynamics of $\phi$, we expand the function $f(\phi,E^\pm)$ around $E^\pm=\bar{E}^\pm$:
    \begin{equation}
        f(\phi,E^\pm)\approx f(\phi,\bar{E}^\pm)+\delta f(\phi,\bar{E}^\pm)
    \end{equation}
    where
    \begin{align}
        \delta f&=\partial_{E^+}f|_{E^+=\bar{E}^+}\ \delta E^++\partial_{E^-}f|_{E^-=\bar{E}^-}\ \delta E^-\\
        &=\sqrt{2\epsilon\bar{E}^+} \ f_1^+(\phi) \tilde{\eta}_{E^+}+\sqrt{2\epsilon\bar{E}^-} \ f_1^-(\phi) \tilde{\eta}_{E^-}
    \end{align}
    and we defined $f_1^+(\phi)=\partial_{E^+}f(\phi,E^\pm)|_{E^\pm=\bar{E}^\pm}$, $f_1^-(\phi)=\partial_{E^-}f(\phi,E^\pm)|_{E^\pm=\bar{E}^\pm}$. 
    At  leading order, once we neglect cross-correlations between noise terms in the dynamics of molecules and enzymes, we can approximate $g(\phi,E^\pm) \approx g(\phi,\bar{E}^\pm)$.
    The equation for $\phi$ can be rewritten as
        \begin{widetext}

    \begin{equation}
        \partial_s \phi=f(\phi,\bar{E}^\pm)+\xi
    \end{equation}
    with
    \begin{align}
        \langle\xi(s,\vec{x})\xi(s',\vec{x}')\rangle&=
        \left\{g(\phi,\bar{E}^\pm) 
        \nonumber 
        +2\epsilon \left[\bar{E}^+f_1^+(\phi)^2+\bar{E}^- f_1^-(\phi)^2\right]
        \right\}
        \delta(\vec{x}-\vec{x}')\delta(s-s')
    \end{align}
    Switching back to the original timescale, the equation for the slow variable $\phi$ reads
    \begin{align}
        \partial_t\phi&=D\nabla^2\phi+A+\sqrt{B} \ \eta
        \end{align}
    with $A$ and $B$ given by
    \begin{align}
        A&=
    \frac{2k_\mathrm{c}\bar{E}^+(c-\phi)}{2\gamma K_\mathrm{m}^++c-\phi}
    +    k^+_\mathrm{b}
    (c-\phi)-\frac{2k_\mathrm{c}\bar{E}^-(c+\phi)}{2\gamma K_\mathrm{m}^-+c+\phi}
        -k^-_\mathrm{b}
        (c+\phi)
    \\
    \label{eq:Bafter_enslavement}
    B&=
    \frac{4k_\mathrm{c}\bar{E}^+(c-\phi)}{2\gamma K_\mathrm{m}^++c-\phi}\left(1+2\epsilon\frac{c-\phi}{2\gamma K_\mathrm{m}^++c-\phi}\right)
    +\frac{4k_\mathrm{c}\bar{E}^-(c+\phi)}{2\gamma K_\mathrm{m}^-+c+\phi}\left(1+2\epsilon\frac{c+\phi}{2\gamma K_\mathrm{m}^-+c+\phi}\right)+    2k^+_\mathrm{b}(c-\phi)  +2
        k^-_\mathrm{b}
        (c+\phi) 
    \end{align}
            
    \end{widetext}
    and $\eta$ a zero-mean noise with correlation
    \begin{equation*}
        \langle\eta(\vec{x},t)\eta(\vec{x}',t')\rangle=\delta(\vec{x}-\vec{x}')\delta(t-t')
    \end{equation*}
    As a result of the adiabatic elimination of the enzyme degrees of freedom, the noise amplitude in Eq.~(\ref{eq:Bafter_enslavement}) now contains additional $\mathcal{O}(\epsilon)$ terms that are not present in the intrinsic noise of the field $\phi$ [cf. Eq.~(\ref{eq:langevin_phi_raw})].
    These corrections have a functional form similar to the intrinsic noise of $\phi$ and thus effectively result in a small renormalization of the noise strength. Since simulations are performed in a parameter regime where $\epsilon\lesssim1$ these corrections are neglected hereafter to maintain analytical tractability.

    In the expression for the stationary concentration of the enzymes $\bar{E}^\pm=\frac{k_\mathrm{a}^\pm E_\mathrm{f}^\pm (c\pm\phi)}{2k_\mathrm{d}^\pm}$, in the fast-shuttling regime we can substitute the cytosolic concentration $E_\mathrm{f}^\pm$ with its stationary value, $E_\mathrm{f}^\pm=\frac{2\gamma K_\mathrm{d}^\pm E_\mathrm{tot}^\pm}{2\gamma K_\mathrm{d}^\pm+c\pm\langle\phi \rangle_\mathrm{s}}$, where $K^\pm_\mathrm{d}=k_\mathrm{d}^\pm/k_\mathrm{a}^\pm$.
        
With these assumptions, we reduced
the system's description to a single stochastic equation for the order parameter $\phi$, given by Eq.~(\ref{eq:phi}) in the main text.
For convenience, we provide a summary of the symbols used for the relevant fields and parameters of the theory in Tables~\ref{tab:fields} and~\ref{tab:parameters}.

\section{Phase diagram}
\label{app:phase_diag_details}
The
effective catalytic
ratio
$\rho=k_\mathrm{e}^+/k_\mathrm{e}^-$
can be 
determined 
as a
function of
the field average~$\langle \phi \rangle_\mathrm{s}$:
\begin{align}
    \label{eq:rho}
\rho&=
\frac{2\gamma K^-_\mathrm{d}+ c - \langle \phi \rangle_\mathrm{s} }{2\gamma K^+_\mathrm{d}+ c + \langle \phi \rangle_\mathrm{s} }\,\rho_0
\end{align}
where 
$$\rho_0=\frac{k^+_\mathrm{c} E^+_\mathrm{tot}}{k^-_\mathrm{c} E^-_\mathrm{tot}}$$ 
Since 
the field average is constrained by
$-c\leqslant \langle \phi \rangle_\mathrm{s} \leqslant c$,
the values of $\rho$
must lie in 
the
{physical
region} 
defined by:
\begin{equation}
    \label{eq:phys_strip2}
    \frac{\rho_0}{\rho_+}
    \leqslant \rho \leqslant 
    \frac{\rho_0}{\rho_-}
\end{equation}
where
\begin{eqnarray}
    \rho_+ =
    \frac{K^+_\mathrm{d}+C}{K^-_\mathrm{d}}, \qquad \rho_- =
    \frac{K^+_\mathrm{d}}{K^-_\mathrm{d} +C} 
\end{eqnarray}

\begin{table}[]
\begin{tabular}{p{0.15\linewidth}
p{0.82\linewidth}}
\specialrule{1.0pt}{2pt}{2pt}
\textit{Symbols} & \textit{Fields} \\
\specialrule{1.0pt}{2pt}{2pt}
$E_\mathrm{f}^\pm(t)$ & Concentration of 
free 
$E^\pm$ enzymes in 
the
unstructured reservoir. 
\\ \midrule
$E^\pm(\vec{x},t)$ & Concentration of $E^\pm$ enzymes on the scaffold.
\\ \midrule
$\phi^\pm(\vec{x},t)$ & Concentration of $\phi^\pm$ 
molecules on the scaffold.
\\ \midrule
$\phi(\vec{x},t)$ & 
Order parameter,
representing the local composition difference: 
$\phi^+(\vec{x},t)-\phi^-(\vec{x},t)$.
\\ \midrule
$\langle\phi(\cdot,t)\rangle_\mathrm{s}$ & Spatial average of the field 
$\phi$ over the scaffold.  
\\ 
\specialrule{1.0pt}{2pt}{2pt}
\end{tabular}
\caption{\label{tab:fields}
Concentration fields included in the model.
\vspace{0.3cm}
}
\end{table}
\begin{table}[]
\begin{tabular}{p{0.15\linewidth}
p{0.82\linewidth}}
\specialrule{1.0pt}{2pt}{2pt}
\textit{Symbols} & \textit{Parameters} \\\specialrule{1.0pt}{2pt}{2pt}
$k_\mathrm{c}^\pm$ & Rate of
 $\mp\to\pm$
conversion. 
\\ \midrule
$K_\mathrm{m}^\pm$ & Michaelis-Menten constant for 
$\mp\to\pm$
conversion. 
\\ \midrule
$k_\mathrm{e}^\pm$ & Effective rate of
$\mp\to\pm$ 
conversion,
accounting for 
enzyme depletion 
from the reservoir.
\\ \midrule
$k_\mathrm{b}^\pm$ & Rate of $\mp\to\pm$  basal conversion. 
\\ \midrule
$k_\mathrm{a}^\pm$ & Rate of
$E^\pm$
binding
to the scaffold.
\\ \midrule
$k_\mathrm{d}^\pm$ & 
Rate of $E^\pm$  unbinding 
from the scaffold. 
\\ \midrule
 $K_\mathrm{d}^\pm$ & 
 Dissociation constant,
 defined as 
 $k_\mathrm{d}^\pm/k_\mathrm{a}^\pm$.
 \\ \midrule
$D$ & Macroscopic 
diffusivitiy on the scaffold. 
\\ \midrule
$\gamma$ & Capacity ratio. 
\\ \midrule
$c$ & 
Total concentration 
$\phi^++\phi^-$. 
\\ \midrule
$E_\mathrm{tot}^\pm$ & Total concentration of $E^\pm$ enzymes.\\
\specialrule{1.0pt}{2pt}{2pt}
\end{tabular}
\caption{\label{tab:parameters}
Model parameters.
}
\end{table}

The physically accessible values of $\rho$
are thus controlled by the rates of enzyme shuttling 
to and from
the cytosol.
At each 
instant
in time, 
the system's
state is 
characterized by a value of $\rho$ inside the physical 
region 
bounded by the solid curves 
in Fig.~\ref{fig:dynamical_pd} (corresponding to $\langle \phi \rangle_\mathrm{s}=\pm c$). 
The parameter $\rho_0$ controls 
the center of this
region,
while
$K_\mathrm{d}/C$ controls 
its width.
Let us assume
now 
that $\rho>1$, 
so that the
$+$ phase is favored. 
The average value of the field $\langle \phi \rangle$ will slowly increase in time due to the expansion of the $+$ phase. This 
increase, in turn,
leads to a decrease in
$\rho$ [see Eq.~(\ref{eq:rho})].
By
the same reasoning,
we find
that 
for $\rho<1$,
the dynamics causes $\rho$ to increase.
In other words, the global feedback mechanism 
always
drives the system towards $\rho=1$ and, if this state is accessible, the system will 
asymptotically reach~it.

From this dynamical picture,
it is
possible to 
extract
information 
that is
independent of the system's specific
trajectory.
From Eq.~(\ref{eq:phys_strip}), it follows that the phase-coexistence line $\rho=1$ is 
reachable,
and therefore, the system can relax in the steady state of
phase coexistence,
if
the condition
\begin{equation}
    \rho_- \leqslant \rho_0 \leqslant \rho_+
\end{equation}
is satisfied.
This allows 
us
to  
explicitly 
draw
the steady-state phase diagram of the system,
shown
in Fig.~\ref{fig:pd_new}(a).
The width of the phase coexistence region on a logarithmic scale is characterized by the ratio:
\begin{equation}
    \frac
{\rho_+}{\rho_-}=
\left(1 + \frac{C}{K^+_\mathrm{d}} \right) \left( 1+ \frac{C}{K^-_\mathrm{d}}\right)
\label{eq:width_pc}
\end{equation}
The region of parameters for which phase coexistence is observed is thus controlled 
by the dissociation constants.

Inverting 
Eq.~(\ref{eq:rho}), 
we
obtain
the
average value of the field $\langle \phi \rangle_\mathrm{s}$ at steady state:
\begin{align}
    \label{eq:phi_eq_1}
    \langle \phi \rangle_\mathrm{s} = 
    \frac{\left(1+ 2 
       K^-_\mathrm{d}/ 
    {C}\right)\rho_0-\left(1+2  
    K^+_\mathrm{d}/ 
    {C}\right)\rho}{\rho_0+\rho}\,c
\end{align}
and
the membrane fraction occupied by each of the $\pm$ phases:
\begin{equation}
     \langle \phi_{\pm} \rangle_\mathrm{s} 
     =  \frac{ 1 +
     K^d_\mp/
     {C}   - 
     (\rho/\rho_0)^{\pm 1}
      K^d_\pm/ 
      {C}
     }
     {1 +  (\rho/\rho_0)^{\pm 1}}\,c
\end{equation} 
The steady-state, phase-coexistence values [Eq.~(\ref{eq:phi_eq})] are 
obtained by substituting
$\rho= 1$ into the  expressions above.

\section{Polynomial approximation}
\label{sec:polyn}
The effective catalytic potential
(\ref{eq:potential})
for $k_\mathrm{b}^\pm=k_\mathrm{b}$
and $K_\mathrm{m}^\pm=K_\mathrm{m}$
is well approximated by the 
polynomial expression:
\begin{equation}
\tilde{V}(\phi)=\frac{a}{4}(c^2-\phi^2)^2
+a\phi_0(c^2-\frac{1}{3}\phi^2)\phi
+k_\mathrm{b}\phi^2
\label{eq:approx}
\end{equation}
where
\begin{eqnarray}
a&=& 2 \, 
\frac
{k^+_\mathrm{e}+k^-_\mathrm{e}} {c^2}
g\left(K_\mathrm{m}/C\right)
\\
\label{eq:gk}
g(\kappa)&=&1+4\kappa(1+\kappa)\ln
\frac{4\kappa(1+\kappa)}{(1+2\kappa)^2} \\
\phi_0&=&\left(1+2
K_\mathrm{m}/C\right)
\frac{k^+_\mathrm{e}-k^-_\mathrm{e}}{k^+_\mathrm{e}+k^-_\mathrm{e}}\,c
\end{eqnarray}
For $k_\mathrm{b}=0$ and $\phi_0=0$ the approximation (\ref{eq:approx})
has the same critical points and
potential barrier between
the two 
wells
as 
the
original
potential~(\ref{eq:potential}). The function $g(\kappa)$
is monotonically decreasing,
with $g(\kappa)\sim 1+4\kappa \ln(4\kappa)$ for 
small~$\kappa$,
and 
$g(\kappa)\sim 1/(8\kappa^{2})$
for large $\kappa$.

Differentiating the approximated potential (\ref{eq:approx}) 
yields the following approximation for the deterministic drift:
\begin{widetext}
\begin{equation}
    \tilde{A}=2(k^+_\mathrm{e}+k^-_\mathrm{e})g(K_\mathrm{m}/C)(1-\phi^2/c^2)(
    \phi-\phi_0)-2k_\mathrm{b}\phi
    =2(\tilde{R}^+-\tilde{R}^-)
\end{equation}
This expression is given in terms of  
polynomial approximations for the reaction terms:
\begin{eqnarray}
\tilde{R}^\pm &=&
k^\pm_\mathrm{e}g(K_\mathrm{m}/C)(1-\phi^2/c^2)(
c+ 2K_\mathrm{m}/\gamma\pm\phi)\mp \frac{k_\mathrm{b}}{2}\phi
\end{eqnarray}
From these, 
a polynomial approximation $\tilde B$ for the noise term
is readily obtained:
\begin{equation}
    \tilde{B}=
    4(\tilde{R}^++\tilde{R}^-)
    =
4(k^+_\mathrm{e}+k^-_\mathrm{e})g(K_\mathrm{m}/C)(1-\phi^2/c^2)\left(c+2K_\mathrm{m}/\gamma+\frac{k^+_\mathrm{e}-k^-_\mathrm{e}}{k^+_\mathrm{e}+k^-_\mathrm{e}}\phi\right)+4k_\mathrm{b}\phi
\end{equation}

\end{widetext}
\section{Droplet nucleation}
\label{app:nucleation}
We consider the nucleation of a
circular
droplet of the metastable phase within a homogeneous sea of the less-stable phase,
in a $d$-dimensional space. 
We 
adopt a radial droplet ansatz 
$\phi(\vec{x},t)=\phi((r-R(t))/w)$, where $r=|\vec{x}|$,
$R(t)$ is the droplet radius, and 
$w$ 
is
the interfacial width. Along the optimal 
nucleation
path, the 
droplet 
growth is monotonic. 
This makes
the mapping $t \mapsto R(t)$ 
invertible,
allowing us to
use 
$R$ as the evolution parameter
instead of time. 
Substituting 
this ansatz 
into the action (\ref{eq:azione})
gives:
\begin{widetext}
\begin{align}
    S
    &=2
    S_d\int_{0}^{R}\mathrm{d}R'\int_0^\infty\mathrm{d}r\left\{ r^{d-1}\frac{A[\phi]+D\left(\frac{d-1}{w r}\phi'+\frac{1}{w^2}\phi''\right)}{B[\phi]}
    \,
    \frac{
    \phi'
    }{w}
    \right\}
\end{align}
where $S_d$ is the solid angle in $d$ dimensions. 
Switching 
to the variable $z=(r-R')/w$ we get:
\begin{equation}
    S=2 
    S_d\int_{0}^{R}\mathrm{d}R'\int_{-R'/w}^\infty\mathrm{d}z\left\{ \frac{\phi'(zw+R')^{d-1}}{B[\phi]}\left(A[\phi]+\frac{D(d-1)}{(zw+R')w}\phi'+\frac{D}{w^2}\phi''\right)\right\}
\end{equation}
where the profile $\phi$ and its derivatives $\phi',\phi''$ are evaluated at $z$. 
The three terms of the action are computed in the thin-wall approximation, which assumes that
$\phi'$ is sharply peaked at the interface and zero elsewhere. 
The first term~is: 
\begin{align}
&\phantom{=}2
    S_d\int_{0}^{R}\mathrm{d}R'\int_{-R'/w}^\infty\mathrm{d}z\left\{ \frac{\phi'(zw+R')^{d-1}A[\phi]}{B[\phi]}\right\}
    \approx
    2
    S_d\int_{0}^{R}\mathrm{d}R'R'^{d-1}\int_{-R'/w}^\infty \frac{\phi'A[\phi]}{B[\phi]}
    \mathrm{d}z
    \approx
    2\frac{S_d}{d}
    R^d\int_{\phi_\mathrm{s}}^{\phi_\mathrm{m}}\frac{A[\phi]}{B[\phi]}\mathrm{d}\phi
\end{align}
where the extrema $\phi_\mathrm{m}$ and $\phi_\mathrm{s}$ are the values of the field in the metastable and in the stable phase, respectively.
Similarly, the second term reads:
\begin{align}
    &\phantom{=}2
    S_d\int_{0}^{R}\mathrm{d}R' \frac{D(d-1)}{w}\int_{-R'/w}^{\infty}\mathrm{d}z \left\{(zw+R')^{d-2}\frac{(\phi')^2}{B[\phi]}\right\}
    \approx
    2 S_d D R^{d-1}\int_{\phi_\mathrm{s}}^{\phi_\mathrm{m}}\frac{\phi'}{B(\phi)}\mathrm{d}\phi
\end{align}
For the remaining contribution,
integrating by parts and using the 
thin-wall approximation to neglect boundary term, we get:
\begin{align}
   &\phantom{=-}2
    S_d\int_{0}^{R}\mathrm{d}R' D\int_{-R'/w}^{\infty}\mathrm{d}z\left\{\frac{(zw+R')^{d-1}}{w^2B(\phi)}\frac{\mathrm{d}}{\mathrm{d}z}\left[\frac{1}{2}(\phi')^2\right]\right\}\\
    &\approx
    -2
    S_d\int_{0}^{R}\mathrm{d}R'D\int_{-R'/w}^{\infty}\mathrm{d}z \left\{ \frac{\mathrm{d}}{\mathrm{d}z}\left[\frac{(zw+R')^{d-1}}{w^2B(\phi)}\right] \frac{1}{2}(\phi')^2\right\}\\ 
    &\approx
    -S_d D R^{d-1}\int_{\phi_\mathrm{s}}^{\phi_\mathrm{m}}\frac{\phi'}{B(\phi)}\mathrm{d}\phi + S_d D\frac{R^d}{d}\int_{\phi_\mathrm{s}}^{\phi_\mathrm{m}}\frac{\phi'^2 B'(\phi)}{B^2(\phi)}
\end{align}
\end{widetext}
Combining the
above contributions, the action for the formation of a droplet of radius $R$ starting from a metastable homogeneous state is:
\begin{align}
    S&=2
S_d\left[R^{d-1}I_1-\frac{R^d}{d}\left(\Delta \tilde{V} -I_2\right)\right]
\label{eq:finalS}
\end{align}

\begin{table*}[]
\begin{tabular}{
l@{\hspace{9pt}}
r@{\hspace{9pt}}r@{\hspace{9pt}}
r@{\hspace{9pt}}
r@{\hspace{9pt}}r@{\hspace{9pt}}r@{\hspace{9pt}}
r@{\hspace{9pt}}
r@{\hspace{9pt}}l}
\toprule \!\!\!~\  Figure
      & 
      $
      k_\mathrm{c}^+
      $ 
      & 
      $
      k_\mathrm{c}^-
      $ 
      & 
      $
      k_\mathrm{d}^\pm
      $ 
      & 
      $
      k_\mathrm{a}^\pm
      $ 
      & 
      ${\gamma}K_\mathrm{m}^+$ 
      & 
      ${\gamma}K_\mathrm{m}^-$ 
      & 
      $\displaystyle\frac{k_\mathrm{b}^\pm}{k_\mathrm{e}^\mathrm{max}}$ 
      & 
      $\displaystyle\frac{n_\mathrm{tot}^{E^\pm}}{L^2}$ 
      & Initial configuration 
      \\ 
      \midrule
~\ 
\ref{fig:pd_new}(a) 
&   \varyingparamters    &   \varyingparamters    &   \varyingparamters       
&   $
10
$  
&   $1$   &   $1$  &   $10^{-2}$    
& $1$ 
& Random $50\%$ mixture \\ \hline
~\ 
\ref{fig:pd_new}(c) 
&   \varyingparamters    &   \varyingparamters    &   $
10
$  
&   $
10
$     
&   $1$   &   $1$  &   $10^{-2}$  
& $1$ 
& Uniform state\\ \hline
~\ 
\ref{fig:pd_new}(d)
&    $
1
$     &    $
0.27
$     &    $
5
$     &    $
5
$   &   $5$   
&   $10$    
&   $0$  
& $1$ 
& Two domains of $+$ phase\\ \hline
~\ 
\ref{fig:phi_av}
&   \varyingparamters    &   \varyingparamters    &   $
20
$    
&   $
20
$     
&   $10$   &   $10$  &   $10^{-3}$      
& \varyingparamters  
& Random $50\%$ mixture \\ \hline
~\ 
\ref{fig:interface_width}(a) 
&   \varyingparamters    &   \varyingparamters    &   $
10
$  
&   $
10
$   
&   $10$   &   $10$  &   $0$  
& $1$ 
& Single domain of $+$ phase \ \\ \hline
~\ 
\ref{fig:interface_width}(b) 
&    $
1
$     &   $
5
$    &   $
10
$     
&  $
10
$  
&   \varyingparamters   &   \varyingparamters  &   $0$    
& $1$ 
& Single domain of $+$ phase \\ \hline
~\ 
\ref{fig:fluctuations} 
&   $
10
$       
&   $
10
$  
&   $
100
$     &   $
100
$      &   \varyingparamters   &   \varyingparamters &   $0$       
& $20$ 
& Random $50\%$ mixture \\ \hline
~\ 
\ref{fig:nucleation_panel}(a,b) 
&    $
1
$     &    $
1
$     &   \varyingparamters      
&  $
10
$  
&   $10$   &   $10$  &   $0$  
& $1$ 
& Single domain of $+$ phase \\ 
\hline
~\ 
\ref{fig:nucleation_panel}(c) 
&   $
1
$     &    $
1
$     &   \varyingparamters  
&  $
10
$  
&   $10$   &   $10$  &   $10^{-3}$  
& $1$  
& Uniform pure $-$ state 
\\ 
\bottomrule
\end{tabular}
\caption{
\label{table:parvalue}
Simulation parameters. 
An asterisk ($*$) denotes  parameters that are varied in the
corresponding plot. 
Lengths are expressed in units of
$\Omega^{1/2}$,
the timescale is set by $k_D=10^{-1}$,
$n_\mathrm{tot}/L^2=10^2$, and
$ k_\mathrm{e}^\mathrm{max}=\max\{k_\mathrm{e}^+,k_\mathrm{e}^-\}$. 
}
\end{table*}
where the coefficients are defined as:
\begin{align}
\label{eq:coeff1}
    I_1&=\frac{D}{2}\int_{\phi_\mathrm{s}}^{\phi_\mathrm{m}}\frac{\phi'}{B(\phi)}\mathrm{d}\phi\\
    \label{eq:coeff2}
    \Delta \tilde{V}&=-\int_{\phi_\mathrm{s}}^{\phi_\mathrm{m}} \frac{A(\phi)}{B(\phi)}\mathrm{d}\phi\\
    \label{eq:coeff3}
    I_2&=\frac{D}{2}\int_{\phi_\mathrm{s}}^{\phi_\mathrm{m}}\frac{\phi'^2B'(\phi)}{B^2(\phi)}\mathrm{d}\phi
\end{align}
The critical radius
is 
finally found by
maximizing the action (\ref{eq:finalS}) with respect to $R$:
\begin{equation}
    \label{sec:criticalradius}
    R_\mathrm{c}=\frac{(d-1)I_1}{\Delta \tilde{V} -I_2}
\end{equation}
It is worth noticing here that in the limit of additive noise (i.e., $B=\mathrm{const}$), this result reduces to the well-known expression from classical nucleation theory.

The action in Eq.~(\ref{eq:finalS}) is valid to describe nucleation event, while the dynamics of a domain of size $\gg R_c$ is mainly deterministic.
The integrals in (\ref{eq:coeff1})--(\ref{eq:coeff3}) can be evaluated 
for 
$K_\mathrm{m}^\pm=K_\mathrm{m}$
and
$k_\mathrm{b}^\pm=k_\mathrm{b}$
using
the
polynomial approximations for the drift and noise term
provided in App.~\ref{sec:polyn},
and the fact that
near phase coexistence, $\phi'\simeq 
{\left[\frac{2}{D}V(\phi)\right]^{1/2}}$. 
As a result, for vanishing anisotropy and basal interconversion
($\phi_0\rightarrow 0$, $k_\mathrm{b}\rightarrow 0$),
we obtain:
 \begin{align}     \label{eq:criticalradius}
     R_\mathrm{c}&=\frac{\sqrt{D(k^+_\mathrm{e}+k^-_\mathrm{e})}}{4\,|k^+_\mathrm{e}-k^-_\mathrm{e}\,|\left(1+\frac{2K_\mathrm{m}}{C}+\frac{2/3}{1+\frac{2K_\mathrm{m}}{C}}\right)\sqrt{g\left(\frac{K_\mathrm{m}}{C}\right)}}
\end{align}
In general, (\ref{eq:criticalradius}) 
is 
an implicit equation for $R_\mathrm{c}$, since the rates $k_\mathrm{e}^\pm$ depend on the average field $\langle\phi\rangle$ through~(\ref{eq:alphas}). However, this dependence is negligible 
when
a single domain of radius $R\ll s$
is present, 
where 
$s$ is
the linear size of the system.
For the theoretical curve in Fig.~\ref{fig:nucleation_panel}, 
we account for this dependence by
assuming 
$\langle\phi\rangle_\mathrm{s}/c=
\frac{2\pi R^2}{s^2}-1$
and  numerically
solving
the coupled system given by
Eqs.~(\ref{eq:alphas}) and (\ref{eq:criticalradius}) for $R_\mathrm{c}$.

\section{Numerical method}
\label{app:numerical}
To efficiently simulate 
the stochastic dynamics described in App.~\ref{app:microscopic},
we developed an ad hoc code 
for a Gillespie-like \cite{Gillespie77} algorithm 
in the Julia language \cite{github}.
For efficient event sampling 
with
dynamically
changing rates, we employ a variant of Algorithms 6 and 7 
from
Ref.~\onlinecite{TZP17}. In short, this technique makes use of a binary tree to store event rates and their partial sums, guaranteeing just $O(\log N)$ time and 2 pseudo random number extractions per Gillespie jump, where~$N$ is the number of possible events. This scheme allows to efficiently implement reaction and diffusion rates that depend on the number of molecules in the reservoir (in addition to the number of molecules at each site), even when a change in the reservoir affects an extensive number of event rates.

For all simulations, we consider a system of linear size $L=100$ and $\gamma=10$, which 
corresponds to a
reservoir volume of $L^2\gamma$. 
Parameters used in the simulations are reported in Table \ref{table:parvalue}.
\section{Multiple species\label{sec:multispecies}}
\begin{figure}[b]
\includegraphics[width=0.19\columnwidth]{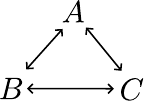}
    \includegraphics[width=0.19\columnwidth]{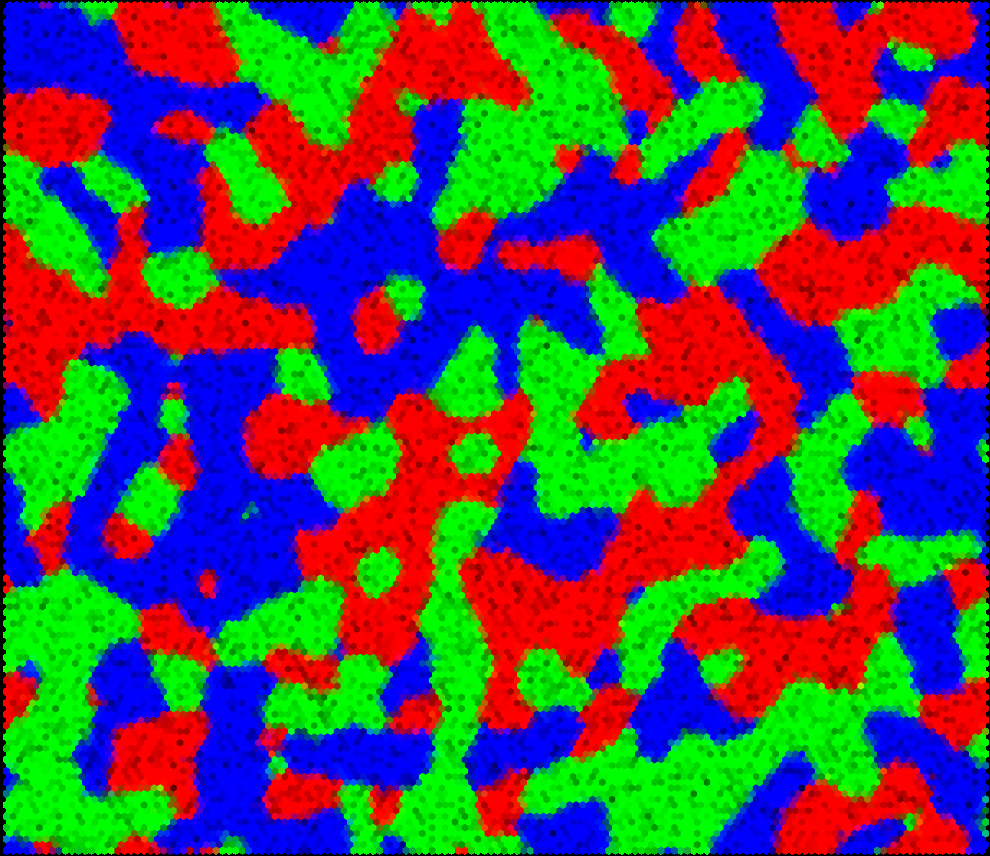}~\includegraphics[width=0.19\columnwidth]{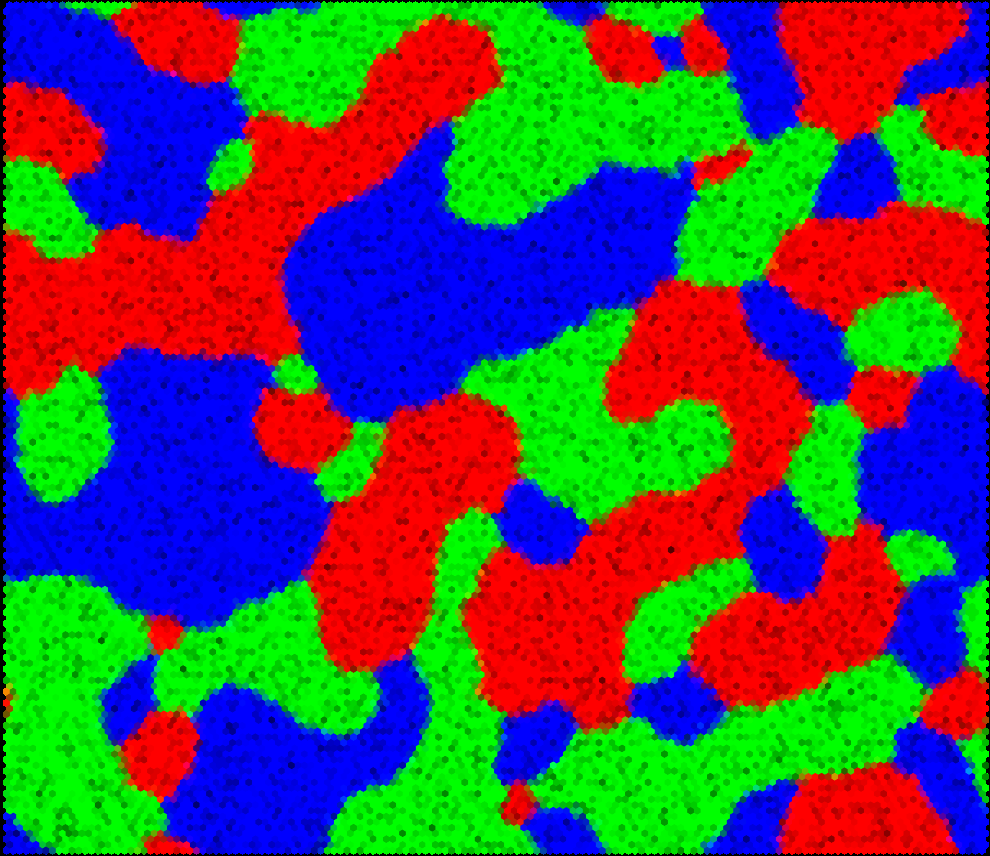}~\includegraphics[width=0.19\columnwidth]{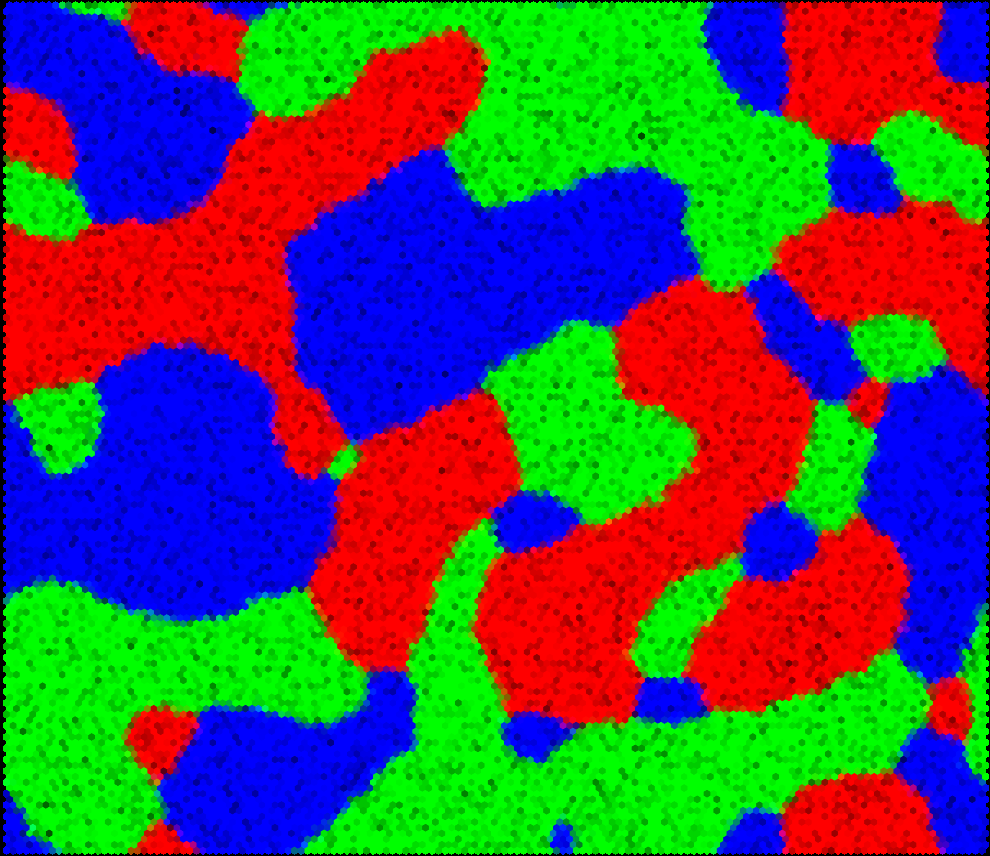}~\includegraphics[width=0.19\columnwidth]{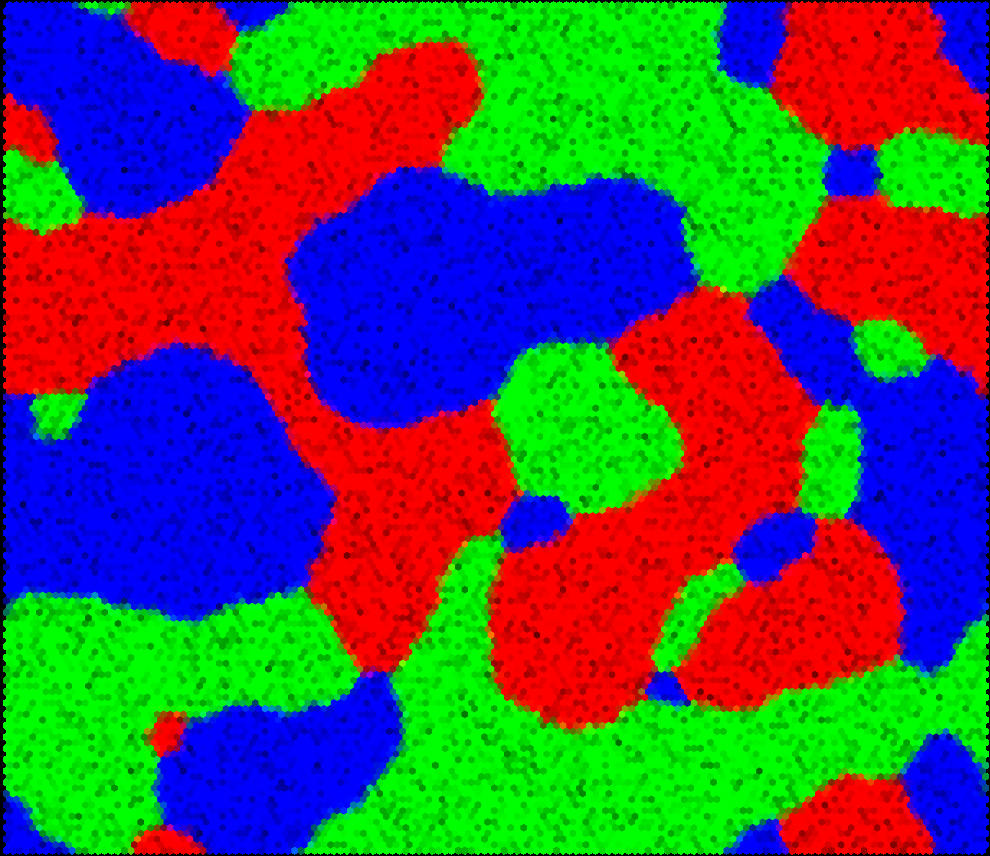}
\includegraphics[width=0.19\columnwidth]{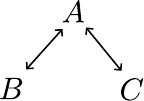}
    \includegraphics[width=0.19\columnwidth]{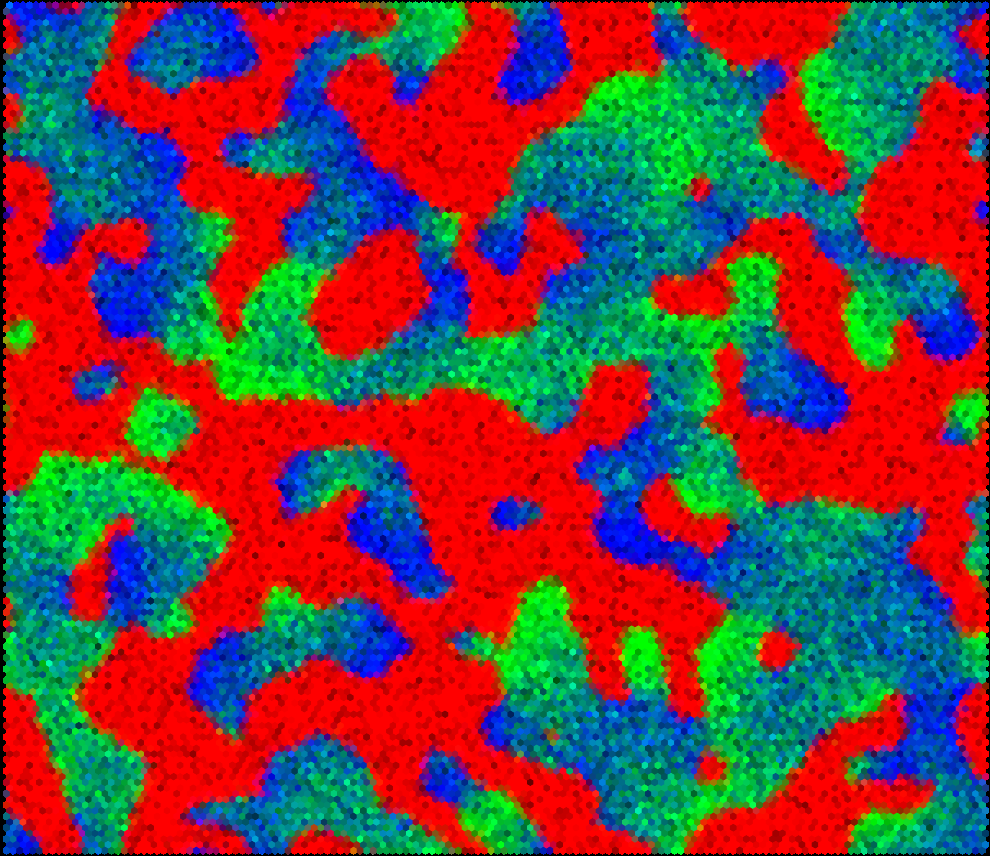}~\includegraphics[width=0.19\columnwidth]{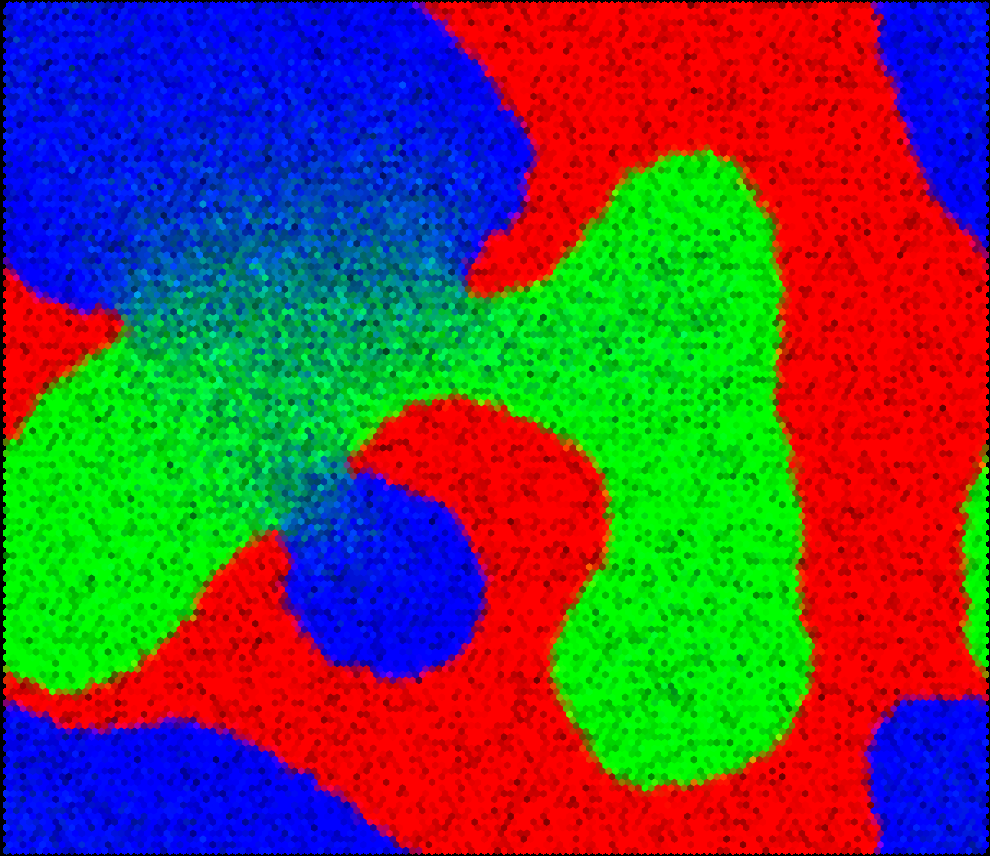}~\includegraphics[width=0.19\columnwidth]{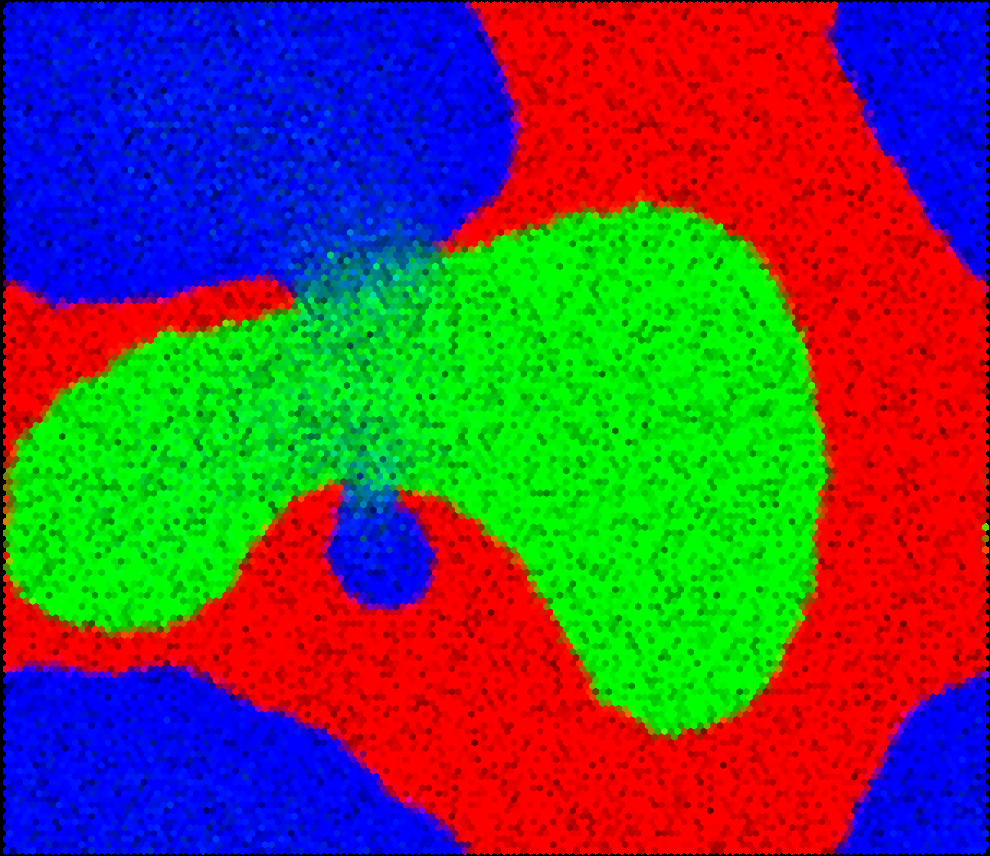}~\includegraphics[width=0.19\columnwidth]{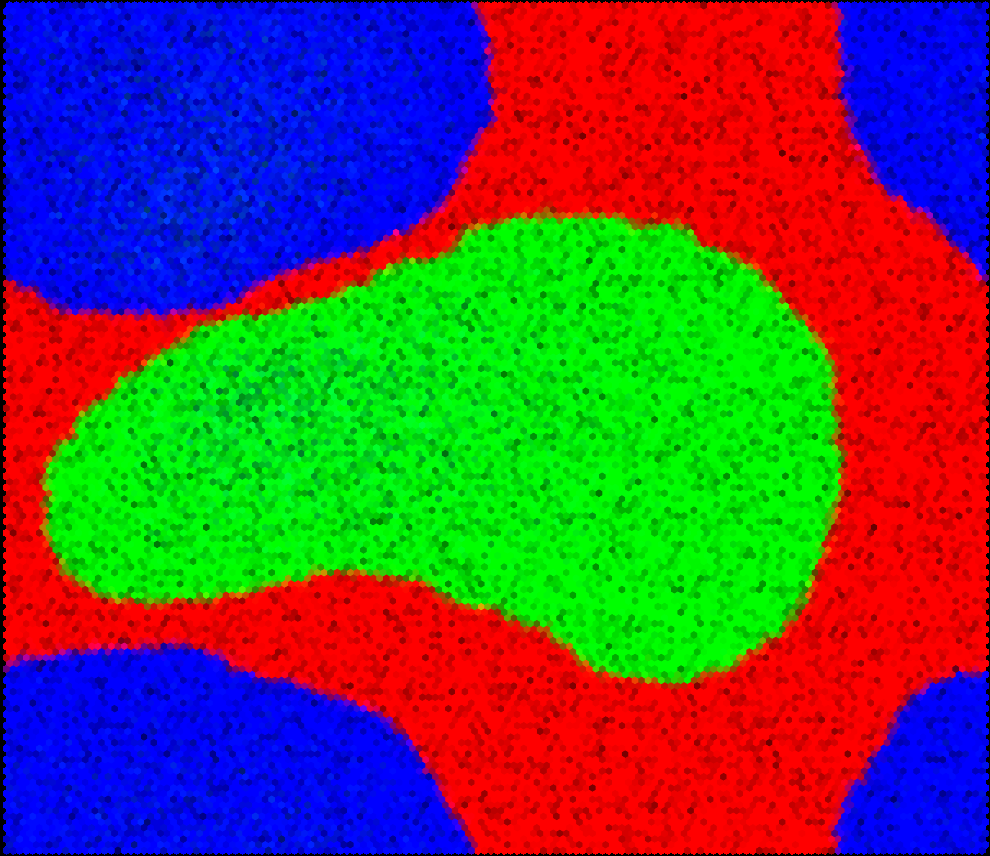}

    \caption{Examples of three-way phase separation in an abstract system with 6 reactions $ E^{xy} +x \to E^{xy}+y $ for $x, y\in\{A =\mathrm{red},B=\mathrm{blue},C=\mathrm{green}\}$, $x\neq y$. The diffusion rate is $k_D = 0.1$ for all species, and $\gamma K_\mathrm{m}=1$ for all reactions. (Top row) Symmetric phase separation, all reaction constants are $k^{xy}_c=1$. (Bottom row) Asymmetric phase separation, $k^{AB}_c=k^{AC}_c=0.8$, $k^{BA}_c=k^{CA}_c=1$ and $k_c^{BC}=k_c^{CB}=0$.\label{fig:3mol} }
\end{figure}

The model can be easily extended 
to 
describe
the activity of several modules working in parallel. As an example, consider the following abstract system of three molecular species $A,B,C$,
with reactions
catalyzed by six enzymes $E^{xy}$,
where
 $x,y\in\{A,B,C\}$ 
and
$ x\neq y$:
\begin{equation*}
    E^{xy}+ x \stackrel{k_c^{xy}}{\longrightarrow} E^{xy}+y 
\end{equation*}
resulting in a three-way 
process of
phase separation. An example of the evolution of such 
a system 
under
symmetric conditions for 
the three molecules is shown in Fig.~\ref{fig:3mol} (top row). When $k_c^{BC}=k_c^{CB}=0$ (resulting in  just four effective reactions), a direct interface between molecules $B$ and $C$ is disfavored, and this interface is 
typically interposed with
molecule $A$ (see Fig.~\ref{fig:3mol}, bottom row).

%

\end{document}